\documentclass{article}
\usepackage{arxiv}
\usepackage[utf8]{inputenc}
\usepackage{float}
\usepackage{subfig}
\usepackage{todonotes}
\usepackage{amsmath}
\usepackage{amssymb}

\usepackage{color}
\usepackage{cleveref}

\title{A Calderon Regularized Symmetric Formulation for the Electroencephalography Forward Problem}

\author{
    John E. Ortiz G. $^{a,b}$\\
    \texttt{john.ortizguzman@polito.it} \\
    \And
    Axelle Pillain $^{a}$ \\
    \texttt{axelle.pillain@imt-atlantique.fr} \\
    \And
    Lyes Rahmouni $^{a,b}$ \\
    \texttt{lyes.rahmouni@polito.it} \\
    \And
    Francesco P. Andriulli $^{b}$ \\
    \texttt{francesco.andriulli@polito.it} \\
}
\date{30 October 2017}
\begin{document}

\maketitle

$^a$ {\footnotesize Computational Electromagnetics Research Laboratory, IMT Atlantique, Brest, France}\\
$^b$ {\footnotesize Politecnico di Torino, Turin, Italy}

\begin{abstract}
The symmetric formulation of the electroencephalography (EEG) forward problem is a well-known and widespread equation thanks to the high level of accuracy that it delivers. However, this equation is first kind in nature and gives rise to ill-conditioned problems when the discretization density or the brain conductivity contrast increases, resulting in numerical instabilities and increasingly slow solutions. This work addresses and solves this problem by proposing a new regularized symmetric formulation. The new scheme is obtained by leveraging on Calderon identities which allow to introduce a dual symmetric equation that, combined with the standard one, results in a second kind operator which is both stable and well-conditioned under all the above mentioned conditions. The new formulation presented here can be easily integrated into existing EEG imaging packages since it can be obtained with the same computational technology required by the standard symmetric formulation. The performance of the new scheme is substantiated by both theoretical developments and numerical results which corroborate the theory and show the practical impact of the new technique.
\end{abstract}


\section{Introduction}

Functional brain imaging based on high-resolution scalp \textcolor{black}{Electroencephalographies} (EEGs) is characterized by \textcolor{black}{a high temporal resolution} and, as such,  it provides an unmatched overview on the underlying brain activity \cite{gavaret2015high, michel2012towards, schmitt2001numerical, phillips1997meg}. This technique relies on the key task, referred to as the EEG inverse problem, of recovering the brain electric current sources responsible for a measured potential at the EEG scalp electrodes \cite{plummer2008eeg, song_eeg_2015}. The EEG inverse problem requires multiple solutions of the EEG forward problem, i.e. the computation of the scalp potential starting from the source currents \cite{grech2008review, hammond2013cortical}. It has been widely studied and reported that the \textcolor{black}{accuracy} of EEG forward problem solvers has a direct impact on  EEG inverse solution procedures \cite{fuchs2001boundary, oostenveld2002validating, acar2013effects,cho2015influence}. For this reason, any advancement of the state of the art in EEG forward solution technologies will have a direct impact on the overall high-resolution EEG imaging process.

When realistic head models are used \cite{acar2013effects,yvert1997systematic, cuffin1996eeg,vatta2010realistic,rullmann2009eeg,aydin2014combining}, the solution of the EEG forward problem can only be obtained numerically. Classical strategies to obtain this numerical solution are the Finite Element Method (FEM), the Finite Difference Method (FDM) or the Boundary Element Method (BEM) \cite{hallez2007review}. Both FDM and FEM leverage on a volume discretization of the considered head model. 
This allows these methods to account for the inhomogeneity and anisotropy of the head's conductivity at the cost, however, of a higher computational demand. Previous works have shown that by using transfer matrices the computation time of the FEM formulations can be reduced \cite{drechsler2009full,genccer2004sensitivity}. The use of these transfer matrices in FEM formulations yield similar computational times as BEM formulations for comparable accuracies. \cite{vorwerk12012comparison}.

When the conductivity of the head is modelled as piecewise homogeneous, BEM can be easily used to compute the solution of the EEG forward problem. In \textcolor{black}{other} words, the main limitation of the BEM formulation resides in its inability to model anisotropies. However, this method has the advantage that it requires only the discretization of the interface between regions with different conductivities \textcolor{black}{\cite{hallez2007review,stenroos2014comparison}}. Several studies (for example \cite{cho2015influence,haueisen1997influence,ramon2004role}) focused on the impact of the head model simplifications in recovering the electric brain sources from the measurement of scalp potential. In particular, when computing the EEG forward problem,  \cite{vorwerk2014guideline,wolters2006influence,marin1998influence} have shown the importance of modelling correctly the skull anisotropy. However, when the anisotropic conductivity values are not known, it can be preferable to model this region as isotropic, as explained in \cite{vorwerk2014guideline}. Moreover, the anisotropic conductivity of the skull is due to its layered structure, \textcolor{black}{a cancellous bone between two compact bones}. This means that when those three layers are available, the skull can accurately be modelled with three isotropic layers instead of one anisotropic layer as \cite{dannhauer2011modeling} shows.

The relevant computational savings which the use of BEM strategies can lead to, explain the attention the technique has received by the community, resulting in a continuous \textcolor{black}{series} of advances \cite{hallez2007review,hamalainen1989realistic, fuchs1998improved, frijns2000improving, ahlfors2010sensitivity, chu2015eeg}. Among them a method, published in \cite{kybic2005common} and referred to as the ``symmetric formulation'',   became quite popular and impacted several EEG based imaging tools \cite{oostenveld2010fieldtrip, gramfort2010openmeeg,tadel2011brainstorm, dalal2011meg}.
The peculiarity of the BEM method  proposed in \cite{kybic2005common} is the quite higher level of accuracy that it can achieve when compared to previously existing schemes. 
%This method has been further improved by \cite{rahmouni2014mixed} who proposed a mixed discretization that abides by the mapping properties of the involved integral operators and results in higher accuracy. 
However these beneficial properties are obtained at the cost of using a first kind formulation (while the majority of standard strategies relies on second kind formulations). The computational consequence of this fact is that, when the ``symmetric formulation'' is discretized to be solved numerically, the condition number of the resulting BEM matrix (the ratio of the largest over the smallest singular value of the matrix) will grow as a function of the discretization density (the number of boundary elements used to discretize the structure) \cite{steinbach2007numerical}. 
 Similarly, a condition number growth is observed in the symmetric formulation also when the conductivity contrast between two regions of the head is increased (a case of practical interest  given that the conductivity of the skull is often modeled with a much smaller value with respect to the conductivity of the brain \cite{acar2013effects, lai2005estimation, zhang2006estimation, Seo-2009}).
In several  cases, especially when handling models issued of high resolution Magnetic Resonance Imaging (MRI) \cite{jirsa2002spatiotemporal}, the solution of the EEG forward problem is obtained iteratively \cite{hallez2007review,kybic2005fast}. A low and stable condition number is desirable  since, on the one hand, the number of iterations of an iterative solver is growing with the condition numbers \cite{axelsson1996iterative} and, on the other hand, the condition number controls the amplification in the solution of any initial error in the sources \cite{axelsson1996iterative}. In other words, the higher the condition number, the longer the time needed to compute the solution, and the less correct the solution will be.

The purpose of this work is to address the ill-conditioning problems of the symmetric formulation. Given the favor that the formulation has found in the community and the fact that it is already implemented in several neuroimaging packages, a particular attention will be devoted to develop a solution strategy that will be conservative, in the sense that will not require the change of previous implementations of the symmetric formulation but will just require the addition of some extra steps to it.  This will be achieved by developing a purely multiplicative preconditioner based on Calderon formulas, i.e. we will design a preconditioning matrix that is spectrally equivalent to the inverse of the symmetric formulation. After left multiplication of this matrix with the symmetric formulation matrix, the resulting linear system will, on the one hand, keep the accuracy the symmetric formulation is well known for and, on the other hand, will provide a stable condition number both when the mesh is refined and when the conductivity contrast between two adjacent domains increases.
\textcolor{black}{The reader should notice that Calder\'on strategies have been successfully applied to the regularization of other integral operators in the context of full-wave vector electromagnetic problems scattered by metallic \cite{christiansen_preconditioner_2002,contopanagos_well-conditioned_2002,adams_numerical_2004,andriulli_multiplicative_2008,stephanson_preconditioned_2009,valdes2011high,zhu_calderon_2011,andriulli_well-conditioned_2013} and penetrable \cite{cools_calderon_2011,niino_calderon_2012,dobbelaere_calderon_2015,gossye2017calderon} objects. The regularization presented here, however, does not automatically follow from none of the strategies above due to the peculiar nature, both in terms of frequency and operator structure/properties, of the symmetric formulation under consideration in this work. It should also be noted that the problem of simulating high contrast dielectric materials has been addressed for high frequency formulations with Calderon strategies both in \cite{niino_calderon_2012} and \cite{gossye2017calderon}. Both of these two clever approaches, designed for the vector case, are not applicable in our scalar scenario where the spectral high-contrast scaling expresses itself as a block scaling issue due to the decoupling of electric and magnetic quantities characterizing a static problem.
}
This paper is organized as follows: Section \ref{Backgr} provides the reader with some necessary background material and notation used in  the following developments. Section \ref{CaldPrec} presents the  new Calderon preconditioner proposed in this work, while Section~\ref{CaldPrec2} focuses on its discretization and on the solution of the preconditioned symmetric formulation system. Section \ref{NumRes} complements the paper's theoretical developments with numerical results which will show the efficiency and effectiveness of the new approach. Partial results from this work has been presented in the conference contribution \cite{guzman2016preconditioning}. 

\section{Background on the EEG Forward Problem}\label{Backgr}

This section will \textcolor{black}{briefly} review the relevant formulations, currently available in literature, \textcolor{black}{used} to solve the EEG forward problem. The treatment will be synthetic and for the sole purpose of setting up the notation. The reader interested in a more profuse treatment should refer, for example to \cite{kybic2005common,montes2014influence} and to references therein.

\subsection{The EEG Problem}
\begin{figure}[!h]
	\centering
	\includegraphics[width=0.4\textwidth]{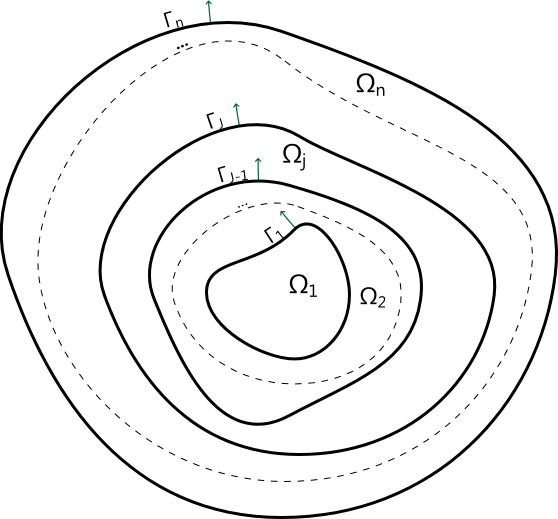}
	\caption{Geometry under consideration.}
	\label{fig:DomainOmega}
\end{figure}

%Let's consider a nested domain $\Omega = \bigcup_{i=1}^{N_{\Omega}} \Omega_{i}$ 
\textcolor{black}{Let $\Omega = \bigcup_{i=1}^{N} \Omega_{i}$ be a nested domain}
with Lipchitz boundaries $\partial \Omega_i =$ %S_i$ such that $S_i =
$\left(\bar{\Omega}_{i-1}\bigcap\bar{\Omega}_i\right) \bigcup \left(\bar{\Omega}_{i}\bigcap\bar{\Omega}_{i+1}\right)$ as in Fig. \ref{fig:DomainOmega}. We denote with $\boldsymbol{n}_i$ the outward going normal to the surface $\Gamma_i$, where $\Gamma_i = \bar{\Omega}_{i}\bigcap\bar{\Omega}_{i+1}$.

Solving the EEG forward problem amounts \textcolor{black}{to computing} the potential $V$ at given electrodes' positions when the active brain current sources are known.
Under quasi-static assumptions and isotropic conductivity, the EEG forward problem reads \cite{sarvas1987basic}:
\begin{equation} \label{fwdPb}
    \sigma\Delta V = \nabla \cdot \boldsymbol j \quad \text{in each $\Omega_i$} 
\end{equation}
where \textcolor{black}{$\Delta=\nabla \cdot\nabla$ is the Laplace operator,} $\sigma$ is the conductivity and $\boldsymbol j$ the current sources. The conductivity is assumed to be piecewise isotropic and homogeneous: in $\Omega_i$, $\sigma = \sigma_i$. \textcolor{black}{In the exterior domain, the conductivity is assumed to be 0.} The current sources, as it is customary in literature  \cite{nunez2006electric}, are assumed to be dipolar in nature. Hence, denoting with $f_i = \nabla \cdot \boldsymbol j$ the electric source in $\Omega_i$, we have $ f_i = q_i \cdot \nabla \delta_{r_i}$ with $q_i$ the electric dipole moment and $\boldsymbol{r}_i$ its position. Furthermore, the symbol $[g]_i = g^--g^+$, will refer to the jump of the function $g$ at the interface $\Gamma_i$, with $g^\mp$ the inner and outer trace of $g$ at $\Gamma_i$ \textcolor{black}{respectively}. Then, the solvability of \eqref{fwdPb} is assured under the following boundary conditions \cite{sarvas1987basic}:
\textcolor{black}{
\begin{subequations}\label{boundCond}
	\begin{equation}\label{contV}
	\left[V\right]_{i} = 0 \; \forall i\leq N
	\end{equation}
	\begin{equation}\label{contDetV}
	\left[\sigma \boldsymbol{n} \cdot \nabla V\right]_{i}=0 \;\forall i\leq N
	\end{equation}
\end{subequations}}
that enforce the continuity of the potential and \textcolor{black}{the current} between the different layers of the domain $\Omega$.

The Green's function \textcolor{black}{associated with} \eqref{fwdPb} reads \cite{steinbach2007numerical}
\begin{equation}
    G(\boldsymbol r,\boldsymbol{r}') = \frac{1}{4\pi |\boldsymbol r-\boldsymbol{r}'|}
\end{equation}
for which we can derive Green's representation theorem using the integral operators \cite{steinbach2007numerical}
\textcolor{black}{
\begin{subequations}\label{Op}
    \begin{align}
    S\Psi (r) &= \int_{\partial_\Omega} G(r,r')\Psi(r') dr' \label{SOp}\\
    D\Phi (r)& = p.v.\int_{\partial_\Omega} \partial_{n'}G(r,r')\Phi(r') dr' \label{DOp}\\
    D^*\Psi (r) &= p.v.\int_{\partial_\Omega} \partial_{n}G(r,r')\Psi(r') dr'. \label{DstarOp}\\
    N\Phi (r) &= f.p.\int_{\partial_\Omega} \partial_{n}\partial_{n'}G(r,r')\Phi(r') dr' \label{NOp}
    \end{align}
\end{subequations}
}
In the above equation \textcolor{black}{and $p.v.$ and $f.p$ stand for Cauchy principal value and Hadamard finite part respectively}. In the following, we will denote with $L_{ij}$ the operator $L$ when $r \in \Gamma_i$ and $r' \in \Gamma_j$ with $L = D, S, N \text{ or } D^*$.

\subsection{The Symmetric Formulation for the EEG Forward Problem}

Several BEM formulations have been proposed to solve the EEG forward problem \cite{hallez2007review, kybic2005common}. Among them, the symmetric formulation \cite{kybic2005common} is quite popular and known for providing high levels of accuracy \cite{gramfort2010openmeeg}.
In solving the EEG forward problem, an efficient strategy is to build the unknown potential $V$ starting from two functions, a function $u$ harmonic in $R^3$ and a function $v$ that takes into account the source term. The starting point of the symmetric formulation is to build $u_i$ in each domain such that $u_i=V-v_i/\sigma_i$ in $\Omega_i$ and $u=-v_i/\sigma_i$ in $R^3\setminus \Omega_i$, with $v_i$ the solution of \eqref{fwdPb} in an unbounded medium: \textcolor{black}{$v_i(r) = \int_{\Omega_i}f_i(r')G(r,r')dr'$}. In this fashion, $u_i$ is harmonic in $R^3\setminus \partial \bar{\Omega_i} = \Gamma_{i-1}\cup \Gamma_i$. Using the boundary conditions \eqref{contV} and \eqref{contDetV} as well as the representation theorem \cite{steinbach2007numerical}, %ref for representation theorem,
two integral equations for the potential and its derivative can be obtained \cite{kybic2005common}. They read:
\begin{multline}\label{InteqPot}
    \sigma_{i+1}^{-1}\left(v_{i+1}\right)_{\Gamma_i} - \sigma_{i}^{-1}\left(v_{i}\right)_{\Gamma_i} = \\
    D_{i,i-1}V_{i-1}-2D_{ii}V_i + D_{i,i+1}V_{i+1}
    -\sigma_{i}^{-1}S_{i,i-1}p_{i-1} + \left(\sigma_{i}^{-1}+\sigma_{i+1}^{-1}\right)S_{ii}p_{i} - \sigma_{i+1}^{-1}S_{i,i+1}p_{i+1}
\end{multline}
\begin{multline}\label{InteqDer}
    \left(\partial_n v_{i+1}\right)_{\Gamma_i} - \left(\partial_n v_{i}\right)_{\Gamma_i} = \\
    \sigma_{i}N_{i,i-1}V_{i-1}-\left(\sigma_{i}+\sigma_{i+1}\right) N_{ii}V_i + \sigma_{i+1}N_{i,i+1}V_{i+1}
    -D^*_{i,i-1}p_{i-1}+2D^*_{ii}p_{i}-D^*_{i,i+1}p_{i+1}
\end{multline}

with $V_{i}$ the potential on the surface $\Gamma_i$ and $p_{i} =  \sigma_i \left[\boldsymbol{n} \cdot \nabla V\right]_{i}$. Equations \eqref{InteqPot} and \eqref{InteqDer} are obtained by applying the boundary conditions on the surface $\Gamma_i$. In a nested domain, it only involves the computation of the operators with functions defined in the surrounding surfaces $\Gamma_{i-1}$, $\Gamma_{i}$ and $\Gamma_{i+1}$.To clarify the ideas, equations \eqref{InteqPot} and \eqref{InteqDer} have been rewritten in matrix form in equation \eqref{Zope}. For a more detailed explanation on the symmetric formulation, the reader is referred to \cite{kybic2005common}.

\begingroup
\begin{equation}\label{Zope}
    {\tiny
    \arraycolsep=1.2pt
    \begin{split}
        &
        \underbrace{\begin{bmatrix}
        	\left(\sigma_1+\sigma_2\right)N_{11} 	& -2D^*_{11} 	&-\sigma_2N_{12}  & D^*_{12} &  &  &  &\\
        	-2D_{11} 	& \left(\sigma_1^{-1}+\sigma_2^{-1}\right)S_{11}  & D_{12}  & -\sigma_2^{-1}S_{12}  &  &  &  &\\
        	-\sigma_2N_{21} 	& D^*_{21} 	&\left(\sigma_2+\sigma_3\right)N_{22}  & -2D^*_{22} & \cdots  &  &  &\\
        	D_{21} & -\sigma_2^{-1}S_{21}  & -2D_{22}  & \left(\sigma_2^{-1}+\sigma_3^{-1}\right)S_{22}  & \cdots & & &\\
        		&  	& \vdots 	& \vdots 	&  \ddots   & & &\\
        	 	& 	&  	&  &   & \left(\sigma_{N-1}+\sigma_N\right)\mathbf{N}_{N-1,N-1} & -2\mathbf{D^*}_{N-1,N-1} & -\sigma_N\mathbf{N}_{N-1,N}\\
        		&  	&   &  &   & -2D_{N-1,N-1} & \left(\sigma_{N-1}^{-1}+\sigma_N^{-1}\right)S_{N-1,N-1} & D_{N-1,N}\\
        		&  	&   &  &   & -\sigma_{N}N_{N,N-1} & D_{N,N-1} & \sigma_{N}N_{N,N}\\
    	\end{bmatrix}}_{Z}\\
    	&
    	\underbrace{\begin{bmatrix}
            V_1\\
            p_1\\
            V_2\\
            p_2\\
            \vdots\\
            p_{N-1}\\
            V_N\\
    	\end{bmatrix}}_{x}
    	=
    	\underbrace{\begin{bmatrix}
        	\left(\partial_n v_{1}\right)_{\Gamma_1} - \left(\partial_n v_{2}\right)_{\Gamma_1}\\
            \sigma_{2}^{-1}\left(v_{2}\right)_{\Gamma_1} - \sigma_{1}^{-1}\left(v_{1}\right)_{\Gamma_1}\\
            \left(\partial_n v_{2}\right)_{\Gamma_2} - \left(\partial_n v_{3}\right)_{\Gamma_2}\\
            \sigma_{3}^{-1}\left(v_{3}\right)_{\Gamma_2} - \sigma_{2}^{-1}\left(v_{2}\right)_{\Gamma_2}\\
            \vdots\\
            \sigma_{N}^{-1}\left(v_{N}\right)_{\Gamma_N-1} - \sigma_{N-1}^{-1}\left(v_{N-1}\right)_{\Gamma_N-1}\\
            \left(\partial_n v_{N}\right)_{\Gamma_N}\\
    	\end{bmatrix}}_{b}
    \end{split}	
    }
\end{equation}

\endgroup

\subsection{Discretization of the Operators} \label{DiscrOp}

The numerical solution of an integral equation is often obtained by using a Boundary Element Method (BEM). Following a well-established strategy, the surface $\Gamma$ is tessellated into $N_t$ triangular cells $t_k$ of area $A_k$ and average length $h$. The set of vertices of the tessellation will be denoted by $\{v_k\}_{k=1}^{N_v}$. Cells and vertices will form a mesh denoted by $M_{\Gamma}$. The number of triangles (respectively, vertices) of the surface $\Gamma_i$ will be denoted $N_{t_i}$ (respectively, $N_{v_i}$ ). To discretize the unknowns and to test the equations the following standard basis functions will be used. The piecewise constant functions in $P_0$ are defined such that, $P_{0k} = 1/A_k$ in $t_k$ and 0 elsewhere. The piecewise linear functions are the set $P_1 = span\{P_{1k}\}_{k=1}^{N_v}$. The support of $P_{1k}$, denoted by $\mu_{P_{1k}}$, is the set of triangles around $v_k$ such that $P_{1k} = 1$ in $v_k$ and $0$ on all other vertices. $P_0$ and $P_1$ functions are shown Fig.~\ref{fig:P0fct} and \ref{fig:P1fct} respectively.

Following \cite{kybic2005common}, on the surface $\Gamma_i$ we respectively discretize the two unknowns $V_i$ and $p_i$ with $P_1$ and $P_0$ basis functions such that 
$V_i = \sum_l^{N_v} a_k P_{1l}$ and $p_i = \sum_l^{N_t} b_l P_{0l}$. \textcolor{black}{In order to} obtain the system matrix, the integral equations \eqref{InteqPot} and \eqref{InteqDer} are tested with $P_0$ and $P_1$ basis functions \textcolor{black}{respectively}. The operator matrices arising are then given by
\begin{subequations}
\begin{align}
        & [\mathbf{D}_{ij}]_{kl} = \int_{t_k} D_{ij}(P_{1l}) \, P_{0k}(r) dr\\
        & [\mathbf{S}_{ij}]_{kl} = \int_{t_k} S_{ij}(P_{0l}) \, P_{0k}(r) dr\\
        & [\mathbf{N}_{ij}]_{kl} = \int_{\mu_{P_{1i}}} N_{ij}(P_{1l}) \, P_{1k}(r) dr\\
        & [\mathbf{D^*}_{ij}]_{kl} = \int_{\mu_{P_{1i}}} D^*_{ij}(P_{0l})\, P_{1k}(r) dr.
    \end{align}
\end{subequations}
Consequently, the system to be solved reads $\mathbf{Z}\mathbf{x} = \mathbf{b}$ with $\mathbf{Z}$ given by
\begingroup
\begin{equation}\label{Zsym}
    {\tiny
    \arraycolsep=1.2pt
	\begin{split}
    	&\mathbf{Z} = \\ 
    	&\begin{bmatrix}
    	\left(\sigma_1+\sigma_2\right)\mathbf{N}_{11} 	& -2\mathbf{D^*}_{11} 	&-\sigma_2\mathbf{N}_{12}  & \mathbf{D^*}_{12} &  &  &  &\\
    	-2\mathbf{D}_{11} 	& \left(\sigma_1^{-1}+\sigma_2^{-1}\right)\mathbf{S}_{11}  & \mathbf{D}_{12}  & -\sigma_2^{-1}\mathbf{S}_{12}  &  &  &  &\\
    	-\sigma_2\mathbf{N}_{21} 	& \mathbf{D^*}_{21} 	&\left(\sigma_2+\sigma_3\right)\mathbf{N}_{22}  & -2\mathbf{D^*}_{22} & \cdots  &  &  &\\
    	\mathbf{D_{21}} & -\sigma_2^{-1}\mathbf{S}_{21}  & -2\mathbf{D}_{22}  & \left(\sigma_2^{-1}+\sigma_3^{-1}\right)\mathbf{S}_{22}  & \cdots & & &\\
    		&  	& \vdots 	& \vdots 	&  \ddots   & & &\\
    	 	& 	&  	&  &   & \left(\sigma_{N-1}+\sigma_N\right)\mathbf{N}_{N-1,N-1} & -2\mathbf{D^*}_{N-1,N-1} & -\sigma_N\mathbf{N}_{N-1,N}\\
    		&  	&   &  &   & -2\mathbf{D}_{N-1,N-1} & \left(\sigma_{N-1}^{-1}+\sigma_N^{-1}\right)\mathbf{S}_{N-1,N-1} & \mathbf{D}_{N-1,N}\\
    		&  	&   &  &   & -\sigma_N\mathbf{N}_{N,N-1} & \mathbf{D}_{N,N-1} & \sigma_N\mathbf{N}_{N,N}\\
    	\end{bmatrix}
	\end{split}
	}
\end{equation}
\endgroup

where
\begin{subequations}
    \begin{align}
        & [\mathbf{x}]_{2l-1} = a_{l}\\
        & [\mathbf{x}]_{2l} = b_l
    \end{align}
\end{subequations}

and where
\begin{subequations}
    \begin{align}
        & [\mathbf{b}]_{2k} = \int_{\mu_k} \left(\sigma_{k+1}^{-1}v_{k+1} - \sigma_{k}^{-1}v_{k} \right) P_{0k} dr\\
        & [\mathbf{b}]_{2k-1} = \int_{t_k} \left(\partial_n v_{k+1} - \partial_n v_{k} \right) P_{1k} dr.
    \end{align}
\end{subequations}

In the following, the  operator \textcolor{black}{associated with} the matrix $\mathbf{Z}$ (obtained by replacing in \eqref{Zsym} $\mathbf{D}$, $\mathbf{S}$, $\mathbf{N}$, and $\mathbf{D^*}$ with ${D}$, ${S}$, ${N}$, and ${D^*}$) will be denoted by $Z$.

\section{A Calderon Preconditioner for the Symmetric Formulation} \label{CaldPrec}

The high accuracy of the symmetric BEM formulation \cite{kybic2005common} has made of it a very popular tool for solving the EEG forward problem. However its system matrix suffers from ill-conditioning that can lead to the non-convergence of the employed iterative solver used to compute the solution \cite{axelsson1996iterative}, \textcolor{black}{which has to be used for high refined meshes where the direct methods are impractical}. Indeed, the operator $S$ is compact \cite{steinbach2007numerical}. This means that its spectrum will accumulate at zero when the mesh is refined and it will therefore have a condition number increasing inversely \textcolor{black}{proportional} to the average mesh length $h$. Moreover, the hypersingular operator $N$ is an unbounded operator \cite{steinbach2007numerical}. 
This implies that its condition number will also grow with $1/h$. Since these operators are the diagonal blocks of the matrix $\mathbf{Z}$ in \eqref{Zsym} and the off-diagonal blocks of the matrix are smoothers \textcolor{black}{\cite{pillain2016line}}, it follows that  the overall conditioning of  $\mathbf{Z}$ will increase when the mesh discretization will increase ($h\to 0$). 

By leveraging on the Calderon identities, it is possible to build a preconditioner for the system matrix $\mathbf{Z}$. The rationale behind our strategy can be understood by considering the continuous operators first. 
The Calderon identities that for our approach read \cite{sauter2011boundary}
\begin{subequations}\label{CaldId}
	\begin{align}
	& S_{ii}N_{ii} = \frac{1}{4} I - D_{ii}^2 \label{Cald1}\\
	& N_{ii}S_{ii} = \frac{1}{4} I - D_{ii}^{*2}\label{Cald2}\\
	& D_{ii}S_{ii} = D^*_{ii}S_{ii} \label{Cald3}\\
	& D^*_{ii}N_{ii} = N_{ii}D^*_{ii} \label{Cald4}
	\end{align}
\end{subequations}
where the symbol $I$ stands for the identity operator. The spectral analysis of \eqref{Cald1} and \eqref{Cald2} shows that the operators $S_{ii}N_{ii}$ and $N_{ii}S_{ii}$ are well conditioned. Indeed, given that $D$ and $D^*$ are compact operators, then $D_{ii}^2$ and $D^{*2}_{ii}$ are also compact operators as a product of two compact operators. Then the spectrum of $D_{ii}^2$ and $D^{*2}_{ii}$ is bounded above and accumulates at zero. However, the presence of the identity operator in \eqref{Cald1} and \eqref{Cald2} guarantees that the spectrum of the operators $S_{ii}N_{ii}$ and $N_{ii}S_{ii}$ will be approaching $1/4$. In other words, $S_{ii}N_{ii}$ and $N_{ii}S_{ii}$ are second kind operators whose spectrum accumulates at $1/4$. This property can be exploited  to build a left preconditioner for the symmetric operator $Z$. 

\textcolor{black}{Before introducing the Calderon preconditioner, we present a regularization matrix for the coefficients of the symmetric operator, which is necessary for the stability of the matrix condition number with respect to the conductivity ratio and for the correct behaviour of the subsequent mesh refinement preconditioner. Indeed, the symmetric operator is unstable when there is a high conductivity contrast between two adjacent domains ,$\Omega_i$ and $\Omega_j$, due to the conductivity factors in the diagonal blocks of the matrix.} These conductivity factors are given by $a_{2l-1,2l-1} = \sigma_{l}+\sigma_{l+1}$ and $a_{2l,2l} = \sigma_{l}^{-1}+\sigma_{l+1}^{-1}$. As a consequence, in the case of high conductivity contrast between two adjacent domains, i.e. asymptotically, when  $\frac{\sigma_i}{\sigma_j} \rightarrow \infty$, that is $\left(\sigma_i+\sigma_j\right) \rightarrow \infty$ when $\sigma_i \rightarrow \infty$ or $\left(\sigma_i^{-1}+\sigma_j^{-1}\right) \rightarrow \infty$ when $\sigma_j \rightarrow 0$, the condition number of the system matrix will grow as a function of the \textcolor{black}{conductivity ratio $CR_{ij}=\frac{\max(\sigma_i,\sigma_j)}{\min(\sigma_i,\sigma_j)}$}. Because of high conductivity contrast between the brain and the skull \cite{lai2005estimation, zhang2006estimation}, this undesirable situation is likely to appear when solving the EEG forward problem. To solve this problem, we will rescale with respect to the conductivity the symmetric operator $Z$ using a diagonal operator $Q$ given by
\begin{equation}
    \textcolor{black}{
    Q=\\
    \begin{bmatrix}
    	\frac{1}{\sqrt{max(\sigma_1,\sigma_2)}} 	& 0 	&0  & 0 & \cdots & 0\\
    	0 	& \sqrt{min(\sigma_1,\sigma_2)} 	&0  & 0 & \cdots & 0\\
    	0 	& 0	& \frac{1}{\sqrt{max(\sigma_2,\sigma_3)}}  & 0 & \cdots & 0\\
    	0 	& 0	& 0  & \sqrt{min(\sigma_2,\sigma_3)} & \cdots & 0\\
    	\vdots 	& \vdots 	& \vdots   & \vdots  & \ddots & \cdots \\
    	0 	& 0	& 0  & 0 & 0 & \frac{1}{\sqrt{\sigma_N}}\\
    \end{bmatrix}
    \label{scalarq}
    }
\end{equation}
We then define a rescaled symmetric operator $Z_q$ as
\begingroup
\arraycolsep=1.2pt
\begin{equation}\label{ZQsym}
    {\tiny
	\begin{split}
    	& Z_q=QZQ = \\ 
    	&\begin{bmatrix}
    	(1+R_{11})N_{11} 	    & -2\sqrt{R_{11}}D^*_{11} 	& -P_2N_{12}  & \sqrt{R_{21}}D^*_{12} &  &  &  &\\
    	-2\sqrt{R_{11}}D_{11}   & (1+R_{11})S_{11}          & \sqrt{R_{12}}D_{12}  & -P^*_2S_{12}  &  &  &  &\\
    	-P_2N_{21} 	            & \sqrt{R_{12}}D^*_{21} 	& (1+R_{22})N_{22}  & -2\sqrt{R_{22}}D^*_{22} & \cdots  &  &  &\\
    	\sqrt{R_{21}}D_{21}     & -P^*_2S_{21}              & -2\sqrt{R_{22}}D_{22}  & (1+R_{22})S_{22}  & \cdots & & &\\
    		&  	& \vdots 	& \vdots 	&  \ddots   & & &\\
    	 	& 	&  	&  &   & (1+R_{N-1,N-1})N_{N-1,N-1} & -2\sqrt{R_{N-1,N-1}}D^*_{N-1,N-1} & -P_NN_{N-1,N}\\
    		&  	&   &  &   & -2\sqrt{R_{N-1,N-1}}D_{N-1,N-1} & (1+R_{N-1,N-1})S_{N-1,N-1} & \sqrt{R_{N-1,N}}D_{N-1,N}\\
    		&  	&   &  &   & -P_NN_{N,N-1} & \sqrt{R_{N-1,N}}D^*_{N,N-1} & N_{N,N}\\
    	\end{bmatrix}
	\end{split}
	}
\end{equation}
\endgroup
with 
\begin{equation}
    R_{ij} = \frac{\min(\sigma_i,\sigma_{i+1})}{\max(\sigma_j,\sigma_{j+1})},\,
    P_i=\frac{\sigma_i}{\sqrt{\max(\sigma_{i-1},\sigma_{i})\max(\sigma_{i},\sigma_{i+1})}},\,
    P^*_i=\frac{\sqrt{\min(\sigma_{i-1},\sigma_{i})\min(\sigma_{i},\sigma_{i+1})}}{\sigma_i}
\end{equation}
and $R_{ij},P_i,P^*_i\leq1$. The asymptotic behaviour of those coefficients is presented in table \ref{table:Coefficients}
, from which we can see that when $CR_{ij}\to \infty$, none of the blocks of $Z_q$ tends to infinity. Furthermore, the diagonal blocks are bounded between 1 and 2. Therefore the instability with respect to high conductivity ratio of the symmetric operator have been solved.
\begin{table*}[h]
    \begin{center}
    \begin{tabular}{ | l | c | c |}
    \hline
    Coefficient & $\sigma_i \to \infty$& $\sigma_i \to 0$ \\ \hline
    $R_{ij}$ & $0$ & $0$ \\ \hline
    $P_{i}$ & $1$ & $0$  \\ \hline
    $P^*_{i}$ & $0$ & $1$  \\ \hline
    \end{tabular}
    \caption{Asymptotic behaviour of the different coefficients of $Z_q$.}
    \label{table:Coefficients}
    \end{center}
\end{table*}

In order to simplify the notation for the following part, we write the regularized symmetric operator as
\begingroup
\arraycolsep=1.3pt
\begin{equation}\label{ZQsym2}
    {\small
    	Z_q= \\
    	\begin{bmatrix}
    	c_{11}N_{11}    & c_{12}D^*_{11} 	& c_{13}N_{12}  & c_{14}D^*_{12} &  &  &  &\\
    	c_{21}D_{11}    & c_{22}S_{11}      & c_{23}D_{12}  & c_{24}S_{12}   &  &  &  &\\
    	c_{31}N_{21}    & c_{32}D^*_{21} 	& c_{33}N_{22}  & c_{34}D^*_{22} & \cdots  &  &  &\\
    	c_{41}D_{21}    & c_{42}S_{21}      & c_{43}D_{22}  & c_{44}S_{22}  & \cdots & & &\\
    		&  	& \vdots 	& \vdots 	&  \ddots   & & &\\
    	 	& 	&  	&  &   & c_{2N-3,2N-3}N_{N-1,N-1} & c_{2N-3,2N-2}D^*_{N-1,N-1}  & c_{2N-3,2N-1}N_{N-1,N}\\
    		&  	&   &  &   & c_{2N-2,2N-3}D_{N-1,N-1} & c_{2N-2,2N-2}S_{N-1,N-1}    & c_{2N-2,2N-1}D_{N-1,N}\\
    		&  	&   &  &   & c_{2N-1,2N-3}N_{N,N-1}     & c_{2N-1,2N-2}D_{N,N-1}        & c_{2N-1,2N-1}N_{N,N}\\
    	\end{bmatrix}
    }
\end{equation}
\endgroup
where the coefficients $c_{ij}$ have the properties
\begin{equation}
    \label{CoefProp}
    c_{ij}=c_{ji},\,c_{2i-1,2i-1}=c_{2i,2i}
\end{equation}

Now we present the Calderon preconditioner for the regularized symmetric operator $Z_q$ which is based on the Calderon identities \eqref{CaldId}. We denote such preconditioning operator by \textcolor{black}{$C_q$}. Its definition is given in \eqref{prec1}, where the constant coefficients $c_{ij}$ are the same as in \eqref{ZQsym2}.
\begingroup
\arraycolsep=1.3pt
\begin{equation}\label{prec1}
    {\small
    C_q=
    \begin{bmatrix}
    	c_{11}S_{11} 	& -c_{12}D_{11} 	&c_{13}S_{12}  & c_{14}D_{12} &  &  &  &\\
    	-c_{21}D^*_{11} 	& c_{22}N_{11}  &c_{23}D^*_{12}  & c_{24}N_{12}  &  &  &  &\\
    	c_{31}S_{21} 	& c_{32}D_{21} 	&c_{33}S_{22}  & -c_{34}D_{22} & \cdots  &  &  &\\
    	c_{41}D^*_{21} & c_{42}N_{21}  & -c_{43}D^*_{22}  & c_{44}N_{22}  & \cdots & & &\\
    		&  	& \vdots 	& \vdots 	&  \ddots   & & &\\
    		& 	&  	&  &   & c_{N-2,N-2}S_{N-1,N-1} & -c_{N-2,N-1}D_{N-1,N-1}  & c_{N-2,N}S_{N-1,N}\\
    		&  	&   &  &   & -c_{N-1,N-2}D^*_{N-1,N-1} & c_{N-1,N-1}N_{N-1,N-1} & c_{N-1,N}D^*_{N-1,N}\\
    		&  	&   &  &   & c_{N,N-2}S_{N,N-1} & c_{N,N-1}D_{N,N-1} & c_{N,N}S_{N,N}\\
    \end{bmatrix}
    }
\end{equation}
\endgroup
We highlight the minus signs present in the blocks $D^*_{ii}$, which are necessary for the compactness of the off diagonal blocks when the preconditioner is applied. A minus sign is added in the $D_{ii}$ blocks as well for symmetry.

Then, as desired, \textcolor{black}{$C_qZ_q$} is a block operator exhibiting the Calderon identities \eqref{Cald1} and \eqref{Cald2} in its diagonal up to the multiplicative factor $c_{ii}^2$ as can be seen in \eqref{CZ1}.
\begin{equation}\label{CZ1}
    C_qZ_q=
    \begin{bmatrix}
        c_{11}^2S_{11}N_{11}+K_{11} & K_{12} & K_{13} & \cdots\\
        K_{21} & c_{22}^2 N_{11}S_{11}+K_{22} & K_{23} & \cdots\\
        K_{31} & K_{32} & c_{33}^2S_{22}N_{22} + K_{33} & \cdots\\
        \vdots & \vdots &  \vdots & \ddots
    \end{bmatrix}
\end{equation}
The terms denoted with $K_{ij}$, contain linear combinations of the compact operators $D_{ij}D_{jk}$, $S_{ij}N_{jk}$, $S_{ij}D^*_{jk}$, $D_{ij}S_{jk}$, $D^*_{ij}N_{jk}$, $N_{ij}D_{jk}$, $N_{ij}S_{jk}$ and $D^*_{ij}D^*_{jk}$. They read
\begin{subequations}
	\label{CZ2}
	\begin{align}
	\begin{split}
	K_{2n-1,2m-1}=&\sum_{i=n-1}^{n+1}\chi_{m}(i)(c_{2n-1,2i-1}c_{2i-1,2m-1}S_{ni}N_{im}+(-1)^{\delta_{ni}}\chi_{i}c_{2n-1,2i}c_{2i,2m-1}D_{ni}D_{im})\\
	&-c_{2n-1,2m-1}S_{nm}N_{nm}\delta_{nm}\\
	\end{split}\\
	\begin{split}
	K_{2n,2m}=&\sum_{i=n-1}^{n+1}\chi_{m}(i)((-1)^{\delta_{ni}}c_{2n,2i-1}c_{2i-1,2m}D^*_{ni}D^*_{im}+\chi_{i}c_{2n,2i}c_{2i,2m}N_{ni}S_{im})\\
		&-c_{2n,2m}N_{nm}S_{nm}\delta_{nm}\\
	\end{split}\\
	&K_{2n-1,2m}=\sum_{i=n-1}^{n+1}\chi_{m}(i)(c_{2n-1,2i-1}c_{2i-1,2m}S_{ni}D^*_{im}+(-1)^{\delta_{ni}}\chi_{i}c_{2n-1,2i}c_{2i,2m}D_{ni}S_{im})\\
	&K_{2n,2m-1}=\sum_{i=n-1}^{n+1}\chi_{m}(i)((-1)^{\delta_{ni}}c_{2n,2i-1}c_{2i-1,2m-1}D^*_{ni}N_{im}+\chi_{i}c_{2n,2i}c_{2i,2m-1}N_{ni}D_{im})
	\end{align}
\end{subequations}
where the symbols $\chi_{i}$, $\chi_{m}(i)$ are given by
\textcolor{black}{
\begin{equation}
    \chi_{m}(i) =
    \left\{
    \begin{array}{ll}
    1  & \mbox{if } |i-m| < 2 \\
    0 & otherwise
    \end{array},
    \right.
\end{equation}
}
\begin{equation}
    \chi_{i} =
    \left\{
    \begin{array}{ll}
    1  & \mbox{if } i < N \\
    0 & otherwise
    \end{array},
    \right.
\end{equation}
and where $\delta_{nm}$ is the Kronecker's delta
\begin{equation}
    \delta_{ij} = \left\{
    \begin{array}{ll}
    1  & \mbox{if } i=j \\
    0 & \mbox{otherwise}
    \end{array}.
    \right.
\end{equation}

As shown previously, the terms $S_{ii}N_{ii}$ and $N_{ii}S_{ii}$ are well conditioned second kind operator matrices, as a consequence the terms $c_{ii}S_{ii}N_{ii}$ $c_{ii}N_{ii}S_{ii}$ will also be well conditioned with respect to the mesh parameter $h$ since $1<c_{ii}<2$. 
\textcolor{black}{
In order to show the compactness of the operators $K_{ij}$, we analyze each operator product in them.
First, in the diagonal blocks $K_{2n-1,2n-1}$ and $K_{2n,2n}$ we have the following products: 
$S_{ni}N_{in}$, $N_{ni}S_{in}$, $D^*_{ni}D^*_{in}$, $D_{ni}D_{in}$. The operators $D$ and $D^*$ are compact, therefore the products $D_{ni}D_{in}$ and $D^*_{ni}D^*_{in}$ are compact as well \cite{sauter2011boundary}.  The products $S_{ni}N_{in}$, $N_{ni}S_{in}$ are present when $i\neq n$, then the operators $N_{in}$ and $N_{ni}$ have a regular kernel due to the lack of singularity. Hence, the products with the compact operators $S_{ni}$ and $S_{in}$ are also compact operators \cite{steinbach2007numerical,gohberg2012basic,abramovich2002invitation}. Overall, in the diagonal block we have the sum of compact operators.
In the off diagonal blocks $K_{2n-1,2m-1}$, $K_{2n,2m}$, when $n \neq m$, we have the products $S_{ni}N_{im}$, $D_{ni}D_{im}$, $D^*_{ni}D^*_{im}$, $N_{ni}S_{im}$. As stated before, the product of the $D^*$ and $D$ are compact. Moreover, in the products $S_{ni}N_{im}$ and $N_{ni}S_{im}$ at least one of the operators has a regular kernel, since $n \neq m$. Hence, these products yield compact operators \cite{gohberg2012basic,abramovich2002invitation}.
Finally, for the off diagonal blocks $K_{2n-1,2m}$, $K_{2n,2m-1}$ we have the following products: $S_{ni}D^*_{im}$ , $D_{ni}S_{im}$, $D^*_{ni}N_{im}$, $N_{ni}D_{im}$. The products $S_{ni}D^*_{im}$ and $D_{ni}S_{im}$ are clearly compact, since all the operators involved are compact. The only not-compact terms are present in the products $D^*_{ni}N_{im}$ and $N_{ni}D_{im}$ when $n=i=m$. However, recalling the Calderon identity \eqref{Cald4} and the properties \eqref{CoefProp}, we have
\begin{equation}
	\label{termDN_ND}
	-c_{2n,2n-1}c_{2n-1,2n-1}D^*_{nn}N_{nn} + c_{2n,2n}c_{2n,2n-1}N_{nn}D^*_{nn} = 0
\end{equation}
which is true for the continues operators. When the operators are discretized, this cancellation is not exact, however the residue tends to zero.  
%there is a cancellation of the principal part(leading order of the operator). With this we conclude that the $K_{ij}$ are compact.
}

\textcolor{black}{
Then, we can write
\begin{equation*}
    C_qZ_q = A+B
\end{equation*}
with
\begin{equation*}\label{matrixA}
	%\begin{align}
	\left\{\begin{aligned}
		&\left[A\right]_{2l-1, 2l-1} &=& c_{2l-1,2l-1}^2S_{ll}N_{ll} \\
		&[A]_{2l, 2l}  &= &c_{2l,2l}^2N_{ll}S_{ll}\\ 
		&[A]_{ij} &= &0 \text{ if } i\neq j
	\end{aligned}
	\right. %\}
\end{equation*}
and $B$ such that $ [B]_{ij} = K_{ij}$, \textcolor{black}{$C_qZ_q$} can be seen as the sum of the well conditioned matrix $A$ and a compact perturbation $B$ (as the operators $K_{ij}$ are compact operators). We can therefore expect the operator \textcolor{black}{$C_qZ_q$} to be well conditioned with respect to the mesh parameter.
}

\textcolor{black}{
For now on, we will refer to the product 
\begin{equation}\label{eq:eqprecon}
	Z_c = C_q Z_q,
\end{equation}
as the Calderon-Symmetric operator, which is well conditioned for both mesh refinement and conductivity ratio between adjacent surfaces.
}

% Finally, for addressing both the mesh parameter and the conductivity related ill-conditionings, the Calderon preconditioning and the rescaling should be performed concurrently. In other words, the preconditioned symmetric operator we propose reads
% \begin{equation}\label{eq:eqprecon}
% 	Z_c = QCQ QZQ,
% \end{equation}
% which is a well conditioned  operator with respect to both the mesh parameter $h$ and the conductivity contrast.

\section{Discretization of the Calderon Preconditioner and Solution of the Preconditioned Symmetric Formulation}\label{CaldPrec2}

In order to solve the preconditioned symmetric integral equation, the proposed multiplicative preconditioner $C_q$ has to be discretized. This discretization should be carried out with care. In fact, the preconditioned operator in \eqref{eq:eqprecon} will contain operator products which will not directly translate into matrix products in the general case. A suitable choice of basis functions should be made to guarantee that this could instead be the case here.
%%%	 Indeed, we must ensure that the matrix multiplication corresponding to the presented operators multiplication can be carried out without perturbing the stability of the proposed preconditioned system.
To fix the ideas, we could consider the discretization of the operator product $N_{11} S_{11}$ appearing in the top-left block of $Z_c$. The matrix $\mathbf{S_{11}}$ is obtained by using \textcolor{black}{test and trial functions} in $ P_0$ while the matrix  $\mathbf{N_{11}}$
is obtained by using \textcolor{black}{test and trial functions} in	
 $P_1$. Yet the number of vertices $N_t$ and the number of cells \textcolor{black}{that defines the dimensions of the space $P_1$ and $P_0$ are different}. As a consequence, the blocks $\mathbf{N_{11}}$ and $\mathbf{S_{11}}$ do not have compatible shapes and cannot be multiplied. %The same reasoning apply to all the blocks multiplication in $C_qZ_q$.
Furthermore, the basis functions used for discretizing $N_{11}$ and $S_{11}$ 
must satisfy appropriate inf-sup conditions with respect to the duality pairing $\langle v,w \rangle : H^{1/2}(\Gamma_1) \times H^{-1/2}(\Gamma_1) \rightarrow \mathbb{R}$ (the reader should refer to \cite{hiptmair2006operator} for further technical details on this topic). This condition enables to get a stable condition number for the Gram matrices, necessarily present in the discretized system as they orthonormalize the two chosen basis and testing functions sets.
To properly take care of this fact, we propose to discretize the preconditioner $C_q$ on the dual mesh $M_{\Omega}^*$ of the standard mesh $M_{\Omega}$ and to leverage on the dual basis functions introduced in \cite{buffa2007dual} on such a mesh. In the dual mesh, each vertex corresponds to a cell of the standard mesh and vice-versa. This means that in $M_{\Omega}^*$ we can build a discrete space in $H^{1/2}(\Gamma)$ which has the same dimension as the discrete space \textcolor{black}{associated with} $H^{-1/2}(\Gamma)$ in $M_{\Omega}$ and vice-versa. Moreover, the dual basis functions introduced by \cite{buffa2007dual} abide by the inf-sup conditions required to obtain stable discrete products \cite{hiptmair2006operator}. As a consequence, the discretization of the preconditioner operator $C_q$ by using these basis functions enables to perform the matrix multiplication \textcolor{black}{associated with} the operator multiplication $C_qZ_q$ in such a  way that the spectral bounds holding for the continuous operator products will translate in well-conditioned matrix products.

\begin{figure}[!h]
	\centering
	\includegraphics[width=0.4\textwidth]{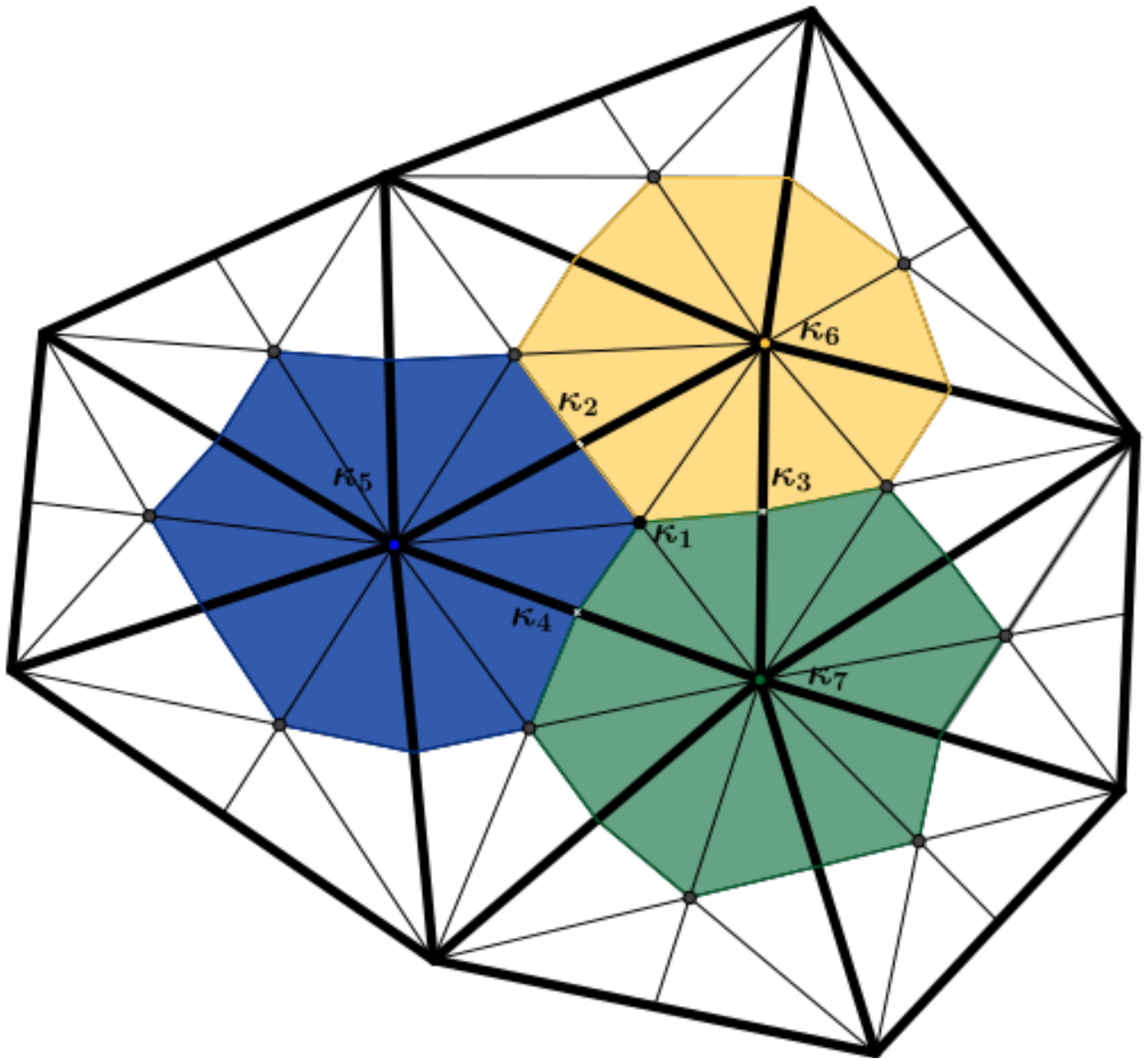}
	\caption{Standard Mesh in bold lines, its barycentric refinement in thin lines. Three cells of the dual mesh are evidenced. The coefficients $\kappa$ are the coefficients used in the linear combination of primal $P_1$ functions to build the dual $\tilde{P}_1$ functions: $\kappa_1 = 1$, $\kappa_i = 1/2$ if $i \in {2,3,4}$ and $\kappa_i = 1/t$ if $i \in {5,6,7}$, with $t$ the number of triangles in $M_{\Omega}$ sharing the corresponding node.}
	\label{fig:barMesh}
\end{figure}

The dual mesh $M_{\Omega}^*$ can be obtained by barycentric refinement of the standard mesh $M_{\Omega}$ by dividing the triangles $t_k$ into six smaller triangles $t_{bk}$ whose edges are built by tracing the medians of the standard triangles $t_k$ \cite{buffa2007dual}. The cells $c_k$ of $M_{\Omega}^*$ are defined as the set of triangle $t_{bk}$ sharing a common vertex $v_k$ of $M_{\Omega}$. The vertices of $M_{\Omega}^*$ are the barycenters $b_k$ of the triangles $t_k$ in $M_{\Omega}$.  The reader should refer to  Fig.~ \ref{fig:barMesh} for an example of such a refinement. 

\begin{figure}[h!]
	\centering
	\subfloat[$P_0$ function]{\label{fig:P0fct}\includegraphics[width=0.3\textwidth]{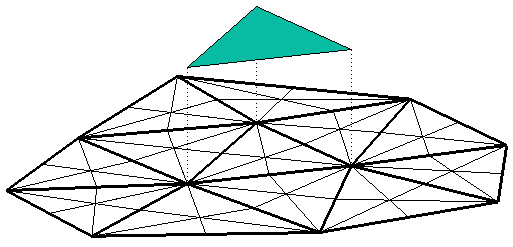}}
	\hspace{5pt}
	\subfloat[$P_1$ function]{\label{fig:P1fct}\includegraphics[width=0.3\textwidth]{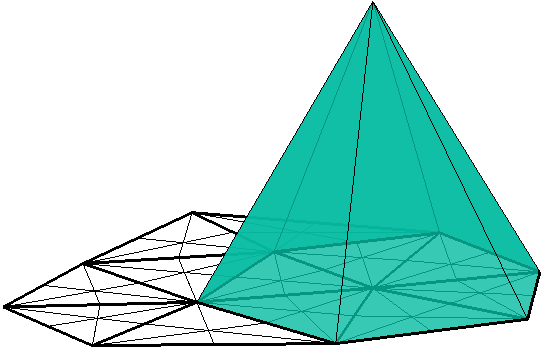}}
	\hspace{5pt}

	\subfloat[$\tilde{P}_0$ function]{\label{fig:dualP0fct}\includegraphics[width=0.3\textwidth]{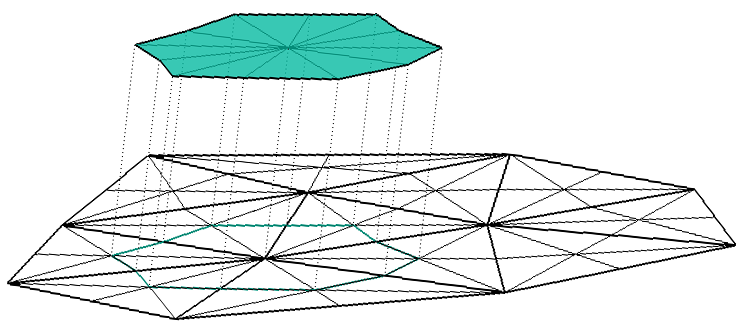}}
	\hspace{5pt}
	\subfloat[$\tilde{P}_1$ function]{\label{fig:dualP1fct}\includegraphics[width=0.3\textwidth]{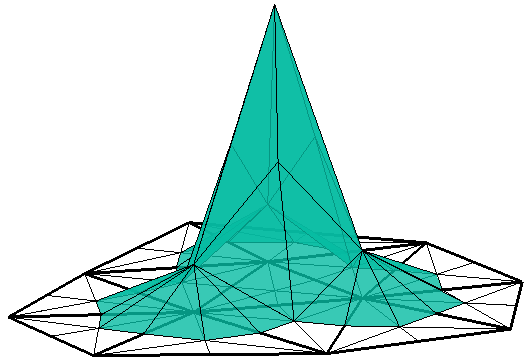}}
	\caption{Basis and testing functions, in the standard mesh (\ref{fig:P0fct}, \ref{fig:P1fct}) and in the dual mesh (\ref{fig:dualP0fct}, \ref{fig:dualP1fct})}
	\label{fig:contour}
\end{figure}

In $M_{\Omega}^*$, we define the dual piecewise linear functions $\tilde{P_1} = span\{\tilde{P}_{1_k}\}_{k=1}^{N_t}$ obtained with a linear combination of $P_1$ functions built on the barycentric refined mesh \cite{buffa2007dual}. A dual piecewise linear function is shown Fig.~\ref{fig:dualP1fct}. The coefficients of the linear combination are shown Fig.~\ref{fig:barMesh}.  The support of $\tilde{P}_{1_i}$ is denoted $\mu_{\tilde{P}_{1_i}}$. The dual piecewise constant functions in $\tilde{P_0}$, denoted $\tilde{P}_{0k}$, are the constant functions equal to $1/A_{c_k}$ on the cell $c_k$, whose area is $A_{c_k}$, of $M_{\Omega}^*$ and equal to zero elsewhere. A dual piecewise constant function is shown \textcolor{black}{in} Fig.~\ref{fig:dualP0fct}. \textcolor{black}{An extended explanation of the dual mesh and dual basis functions is given by \cite{buffa2007dual}.}

Then the operators matrices used to build $\mathbf{C_q}$ are given by:
\begin{subequations}
	\begin{align}
	& [\mathbf{\tilde{D}}_{ij}]_{kl} = \int_{c_k} D_{ij}(\tilde{P}_{1l}) \, \tilde{P}_{0k}(r) dr\\
	& [\mathbf{\tilde{S}}_{ij}]_{kl} = \int_{c_k} S_{ij}(\tilde{P}_{0l}) \tilde{P}_{0k}(r) dr\\
	& [\mathbf{\tilde{N}}_{ij}]_{kl} = \int_{\mu_{\tilde{P}_{1k}}} N_{ij}(\tilde{P}_{1l}) \tilde{P}_{1k}(r) dr\\
	& [\mathbf{\tilde{D}^*}_{ij}]_{kl} = \int_{\mu_{\tilde{P}_{1k}}} D^*_{ij}(\tilde{P}_{0l}) \tilde{P}_{1k}(r) dr.
	\end{align}
\end{subequations}
and the discretized preconditioner $\mathbf{\tilde{C}_q}$ is made explicit as
\begingroup
\arraycolsep=1.3pt
\begin{equation}\label{prec2}
    {\small
    \mathbf{\tilde{C}_q}=\\
    \begin{bmatrix}
        c_{11}\mathbf{\tilde{S}_{11}} 	& c_{12}\mathbf{\tilde{D}_{11}} & c_{13}\mathbf{\tilde{S}_{12}} & c_{14}\mathbf{\tilde{D}_{12}} &  &  &  &\\
        c_{21}\mathbf{\tilde{D}^*_{11}} & c_{22}\mathbf{\tilde{N}_{11}} & c_{23}\mathbf{\tilde{D}^*_{12}} & c_{24}\mathbf{\tilde{N}_{12}} &  &  &  &\\
        c_{31}\mathbf{\tilde{S}_{21}} 	& c_{32}\mathbf{\tilde{D}_{21}}	& c_{33}\mathbf{\tilde{S}_{22}} & c_{34}\mathbf{\tilde{D}_{22}} & \cdots  &  &  &\\
        c_{41}\mathbf{\tilde{D}^*_{21}} 	& c_{42}\mathbf{\tilde{N}_{21}} & c_{43}\mathbf{\tilde{D}^*_{22}} & c_{44}\mathbf{\tilde{N}_{22}} & \cdots & & &\\
        &  	& \vdots 	& \vdots 	&  \ddots   & & &\\
        &  	&   &  &   & c_{N-1,N-2}\mathbf{\tilde{D}^*_{N-1,N-1}} & c_{N-1,N-1}\mathbf{\tilde{N}_{N-1,N-1}} & c_{N-1,N}\mathbf{\tilde{D}^*_{N-1,N}}\\
        &  	&   &  &   & c_{N,N-2}\mathbf{\tilde{S}_{N,N-1}} & c_{N,N-1}\mathbf{\tilde{D}_{N,N-1}} & c_{N,N}\mathbf{\tilde{N}_{N,N}}\\
    \end{bmatrix}
    }
\end{equation}
\endgroup

The final preconditioner is then obtained by introducing the necessary rescaling  to obtain a uniform conditioning with respect to the conductivity profiles (according to the analysis of the previous section). 

\textcolor{black}{In order to perform the multiplication of matrices with two different discretizations, a Gram matrix $\mathbf{G}$ to link them is necessary. This Gram matrix is computed by taking the scalar product between the trial functions of one operator and the test functions of the other operator. Hence, the computation of the Gram matrix does not require the evaluation of any operator. Additionally, it is almost diagonal, therefore the computational cost is very low.  This matrix is obtained as }
% The Gram matrix needed to link the two discretizations is defined as
\begin{subequations}
	\begin{align}
	& [\mathbf{G}_{2i-1}]_{kl} = \int_{\mu_{\tilde{P}_{1k}}} \tilde{(P_{0l})} \, P_{1k}(r) dr \\
	& [\mathbf{G}_{2i}]_{kl} = \int_{t_k} \tilde{(P_{1l})} \, P_{0k}(r) dr.
	\end{align}
\end{subequations}
Finally, the \textcolor{black}{discretization of the Calderon symmetric operator is given by}
\begin{equation}
    \mathbf{Z}_c  = \mathbf{\tilde{C}_q}\mathbf{G}^{-1}\mathbf{Z_q}\
\end{equation}
where $\mathbf{Z_q}=\mathbf{Q}\mathbf{Z_q}\mathbf{Q}$. The solution of the preconditioned symmetric formulation is then obtained by solving the following system
$\mathbf{Z}_c \mathbf{y} = \mathbf{C}_q \mathbf{G}^{-1} \mathbf{Q} \mathbf{b}$ and $\mathbf{x}$ is obtained with $\mathbf{x}=\mathbf{Q}\mathbf{y}$.

Summarizing, the Calderon preconditioning strategy is multiplicative in nature. Its aim is to build a preconditioning operator spectrally equivalent to the inverse of the original operator. Thus, once this operator is built, multiplying the ill-conditioned operator with it yields an operator spectrally equivalent to an identity. The preconditioning operator is built on a dual mesh in order to allow matrix multiplication and stability. Moreover, regularization matrices are added in order to get a condition number independent \textcolor{black}{of} the conductivity ratio. In a nutshell, the steps are:
\begin{enumerate}  
    \item Compute the standard symmetric system matrix $\mathbf{Z}$;
    \item Compute the Calderon preconditioning matrix $\mathbf{\tilde{C}_q}$ on the dual mesh;
    \item Compute the Gram matrices linking the dual and standard discretization, known as Gram matrices $\mathbf{G}$;
    \item Normalize the operator $\mathbf{Z}$ with the regularization matrices $\mathbf{Q}$;
    \item Perform the multiplication $\mathbf{Z}_c  = \mathbf{\tilde{C}_q}\mathbf{G}^{-1}\mathbf{Z_q}$;
    \item The right hand side b must be modified accordingly : compute $\mathbf{b}_c = \mathbf{\tilde{C}_q}\mathbf{G}^{-1}\mathbf{Q}\mathbf{b}$;
    \item Solve the system $\mathbf{Z}_c \mathbf{y} = \mathbf{b}_c$;
    \item Get the solution using $\mathbf{x}=\mathbf{Q}\mathbf{y}$.
\end{enumerate}

\section{Numerical Results}\label{NumRes}

\begin{figure}[h!]
	\centering
	\hspace{1pt}
	\subfloat[Three spheres model\label{SphereModel}]{\includegraphics[width=0.4\textwidth]{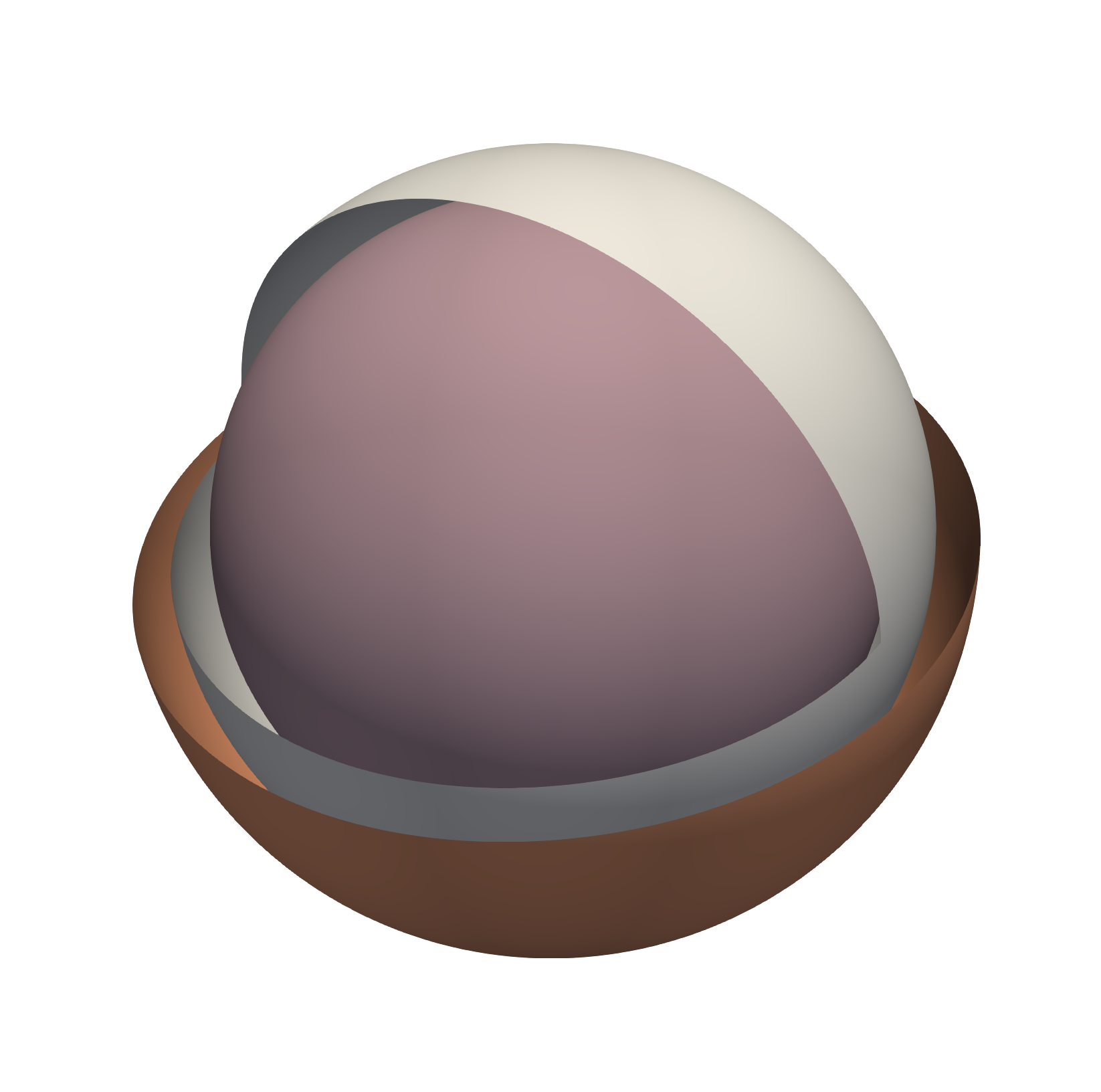}}
	\hspace{1pt}
	\subfloat[MRI-obtained model\label{MRImodel}]{\includegraphics[width=0.4\textwidth]{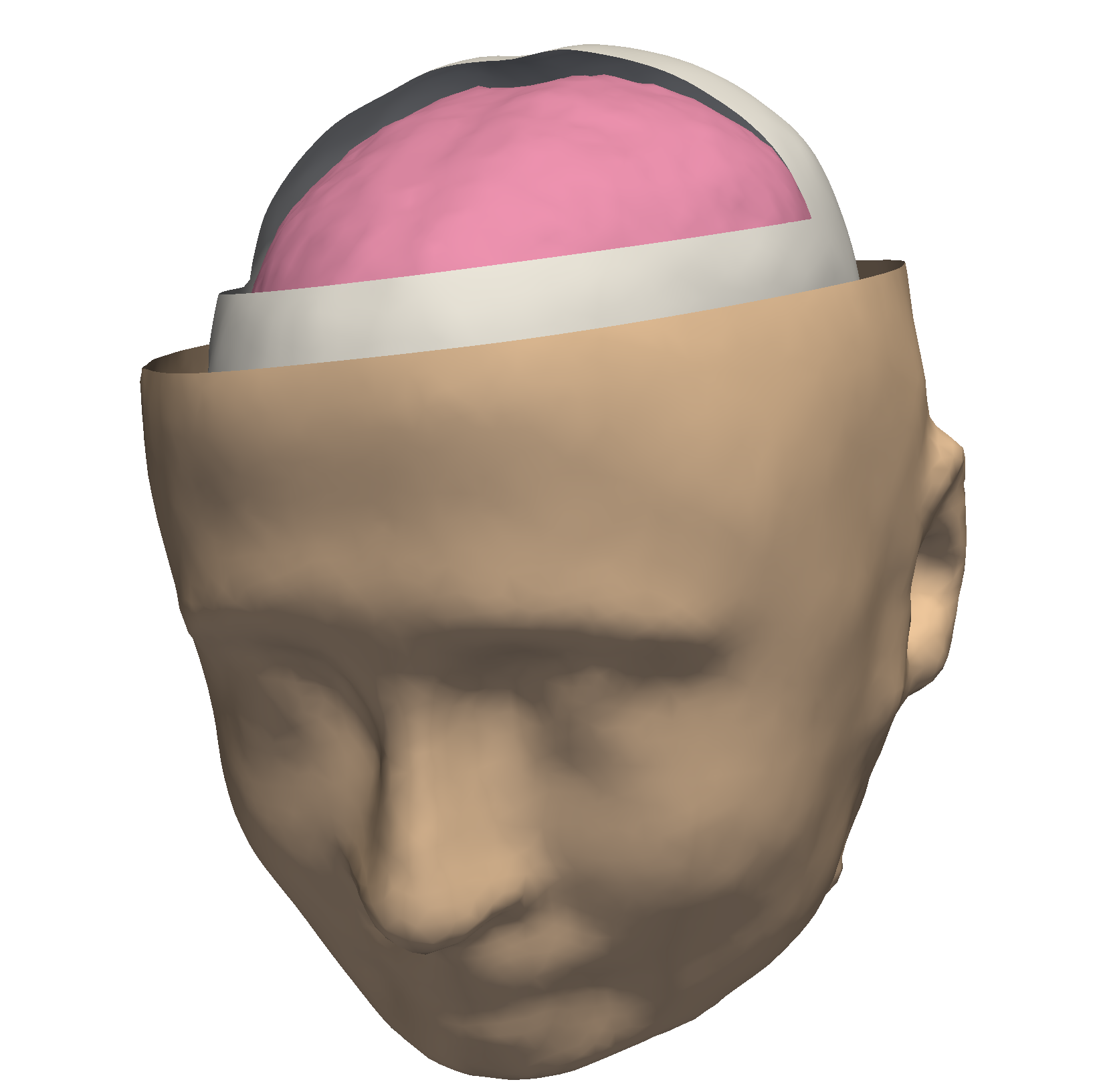}}
	\caption{Head models used for testing the Calderon-Symmetric formulation}
	\label{HeadModels}
\end{figure}

The new Calderon regularized symmetric formulation proposed in this work has been  first tested on the canonical scenario of three homogeneous and concentric spheres of radii 0.8, 0.9, and 1 in normalized units respectively. \textcolor{black}{Such model is shown in Fig.~\ref{SphereModel}}. Indeed, in the case of homogeneous nested spheres, an analytical solution is available as a reference \cite{de1988potential, zhang1995fast}, this solution will be denoted with $V_{ref}$. In these simulations, a single dipole source is placed in (0, 0, 0.5) with a dipole moment of (0, 0, 1) and the \textcolor{black}{normalized} conductivity of the layers is chosen to be 1, 0.0125 and 1 starting from the inner domain, \textcolor{black}{which are the conductivity values used in the standard symmetric formulation in \cite{kybic2005common}}. As a complement to these results, and to validate the new formulation on a real case scenario, the new formulation has been tested also on a realistic head model obtained from MRI data \textcolor{black}{, which is presented in Fig.~\ref{MRImodel}}.

\subsection{Assessments on accuracy and condition number}
\begin{figure}[h!]
	\centering
	\hspace{1pt}
	\subfloat[Accuracy vs Edge\label{fig:accuracy}]{\includegraphics[width=0.4\textwidth]{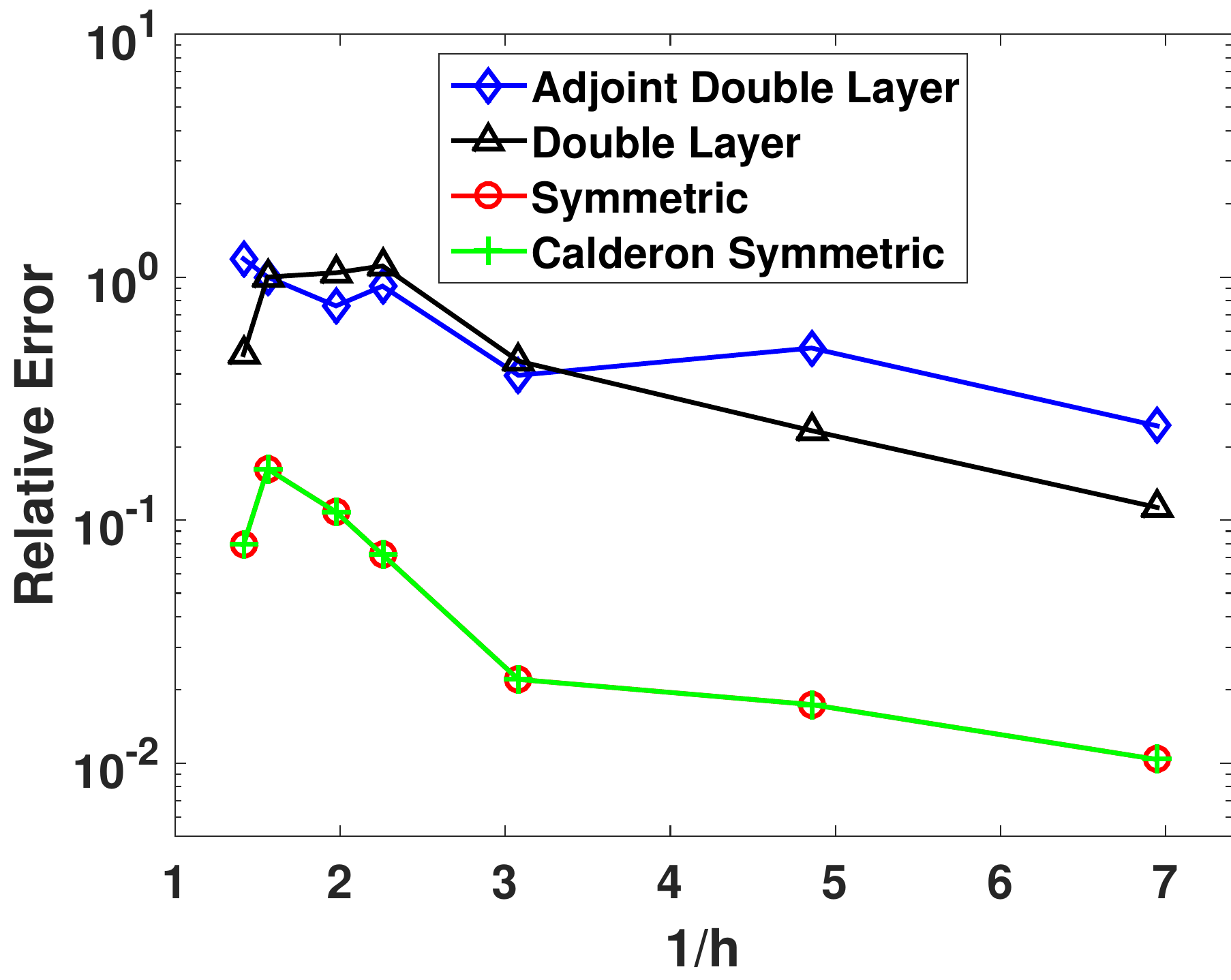}}
	\hspace{1pt}
	\subfloat[Accuracy vs Conductivity Ratio\label{fig:accuracyRatio}]{\includegraphics[width=0.4\textwidth]{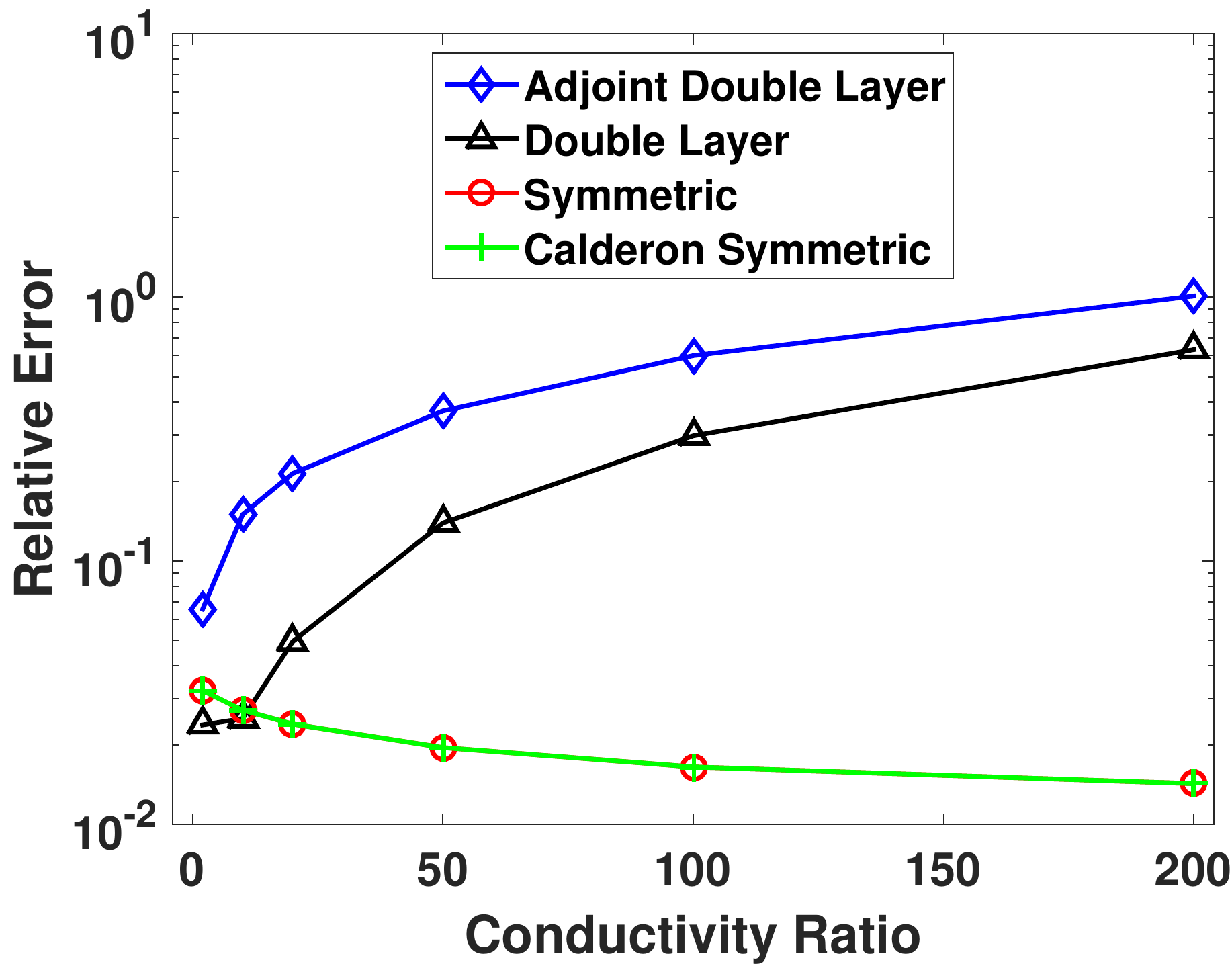}}
	\caption{Accuracy of different boundary integral formulations of the EEG forward problem}
	\label{fig:Acc}
\end{figure}

The first test conducted aimed at verifying that applying the proposed preconditioner to the symmetric formulation does not modify its accuracy. The assessment parameter is the relative error computed as $RE = \frac{||V_{num} - V_{ref}||}{||V_{ref}||}$ where $V_{num}$ refers to the numerical solution. % In Figure~\ref{fig:AccCnd}
%Since a A preconditioner improves the condition of the matrix, and as a consecuence it speeds up the convergence,  mantaining  the same accuracy of the formulation.
In Fig.~\ref{fig:accuracy} \textcolor{black}{and Fig.~\ref{fig:accuracyRatio}} it is shown that the Calderon preconditioned symmetric formulation and the non preconditioned symmetric formulation provide exactly the same accuracy for different mesh refinement levels and \textcolor{black}{different conductivity ratios respectively}. This means that the proposed preconditioner does not alter the accuracy of the initial formulation. \textcolor{black}{These} figures also confirm the higher level of accuracy that the symmetric formulation can reach with respect to two others existing BEM formulations, namely the adjoint double layer formulation and the double layer formulation. 

\begin{figure}[h!]
	\centering
	\hspace{1pt}
	\subfloat[Condition number vs Edge\label{fig:condEdge}]{\includegraphics[width=0.4\textwidth]{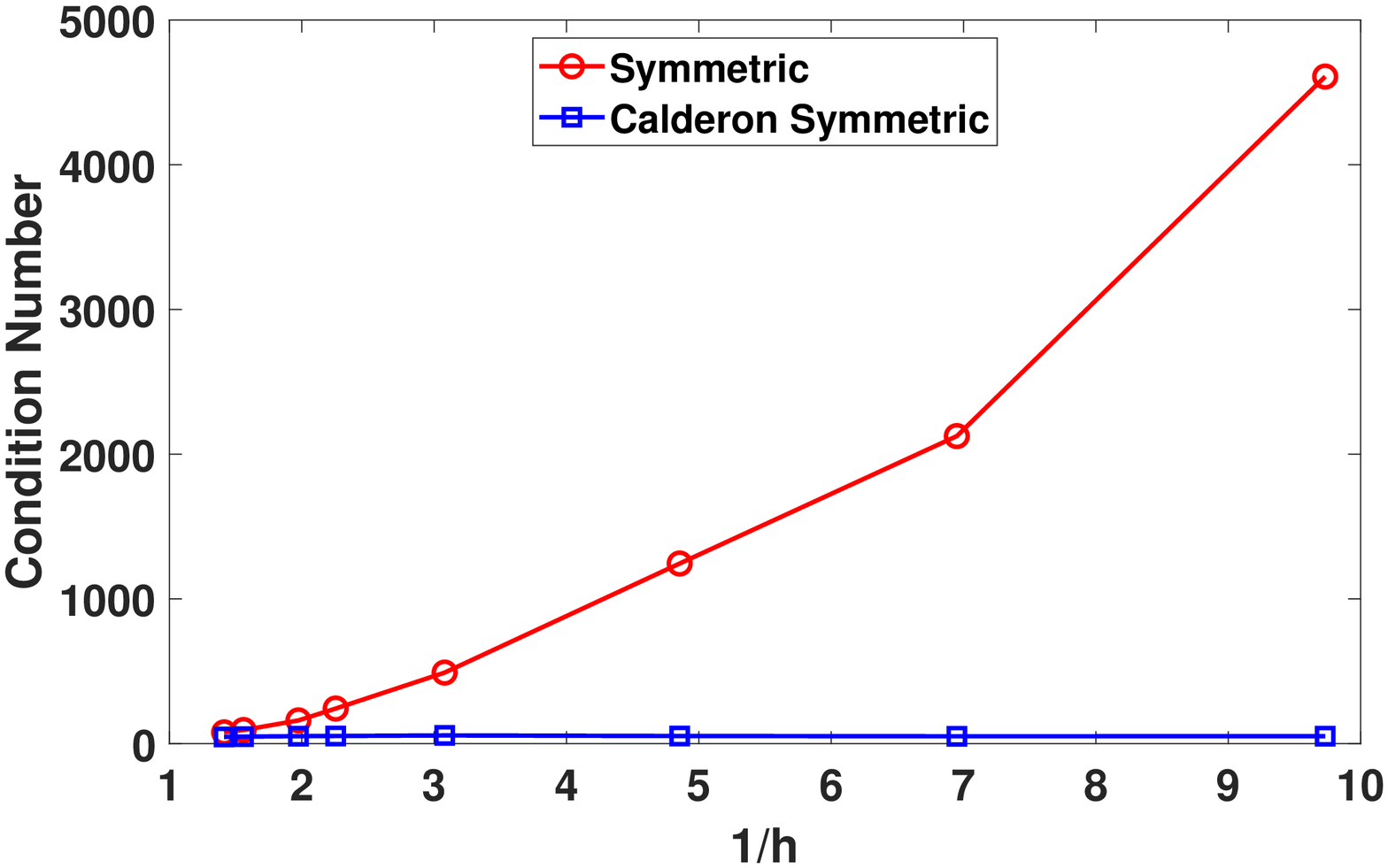}}
	\hspace{1pt}
	\subfloat[Condition number vs Conductivity Ratio\label{fig:condRatio}]{\includegraphics[width=0.4\textwidth]{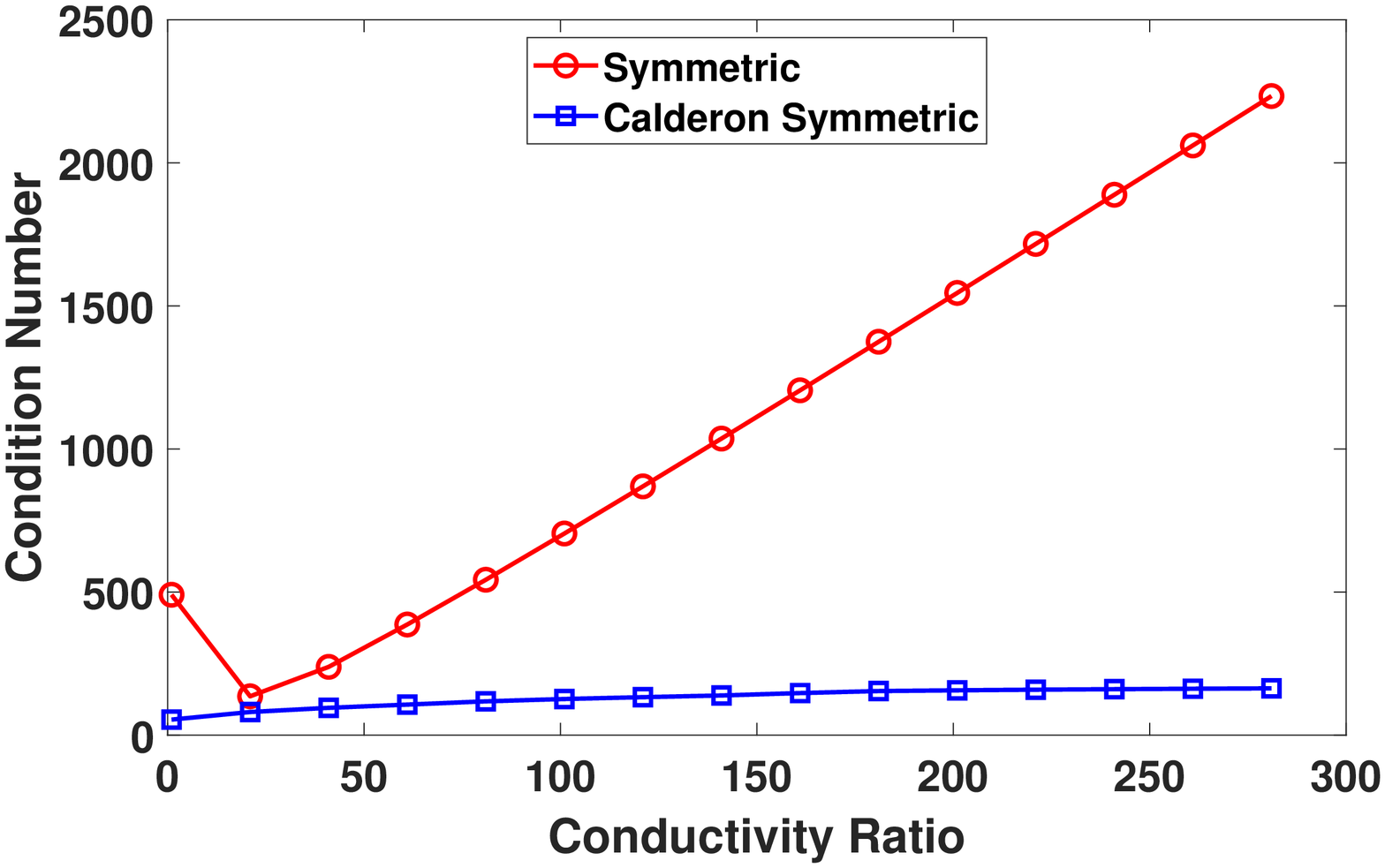}}
	\caption{Condition number of the Symmetric Formulation and Calderon-Symmetric formulation.}
	\label{fig:Cnd}
\end{figure}

In the second test we compared the condition numbers of the preconditioned and the non preconditioned symmetric formulation system matrices . \textcolor{black}{The variation of the condition number with respect to the mesh refinement is shown in Fig.~\ref{fig:condEdge}}. We see that the condition number of the symmetric formulation grows rapidly with the mesh refinement parameter $1/h$ while the condition number of the proposed formulation stays constant as expected by the theory. \textcolor{black}{In Fig.~\ref{fig:condRatio} is presented the condition number when the conductivity ratio increases. Here, we can see that the new formulation has a condition number independent of the conductivity contrast.} 

\begin{figure}[h!]
	\centering
	\hspace{1pt}
	\subfloat[Number of iterations vs Edge\label{fig:iterEdge}]{\includegraphics[width=0.45\textwidth]{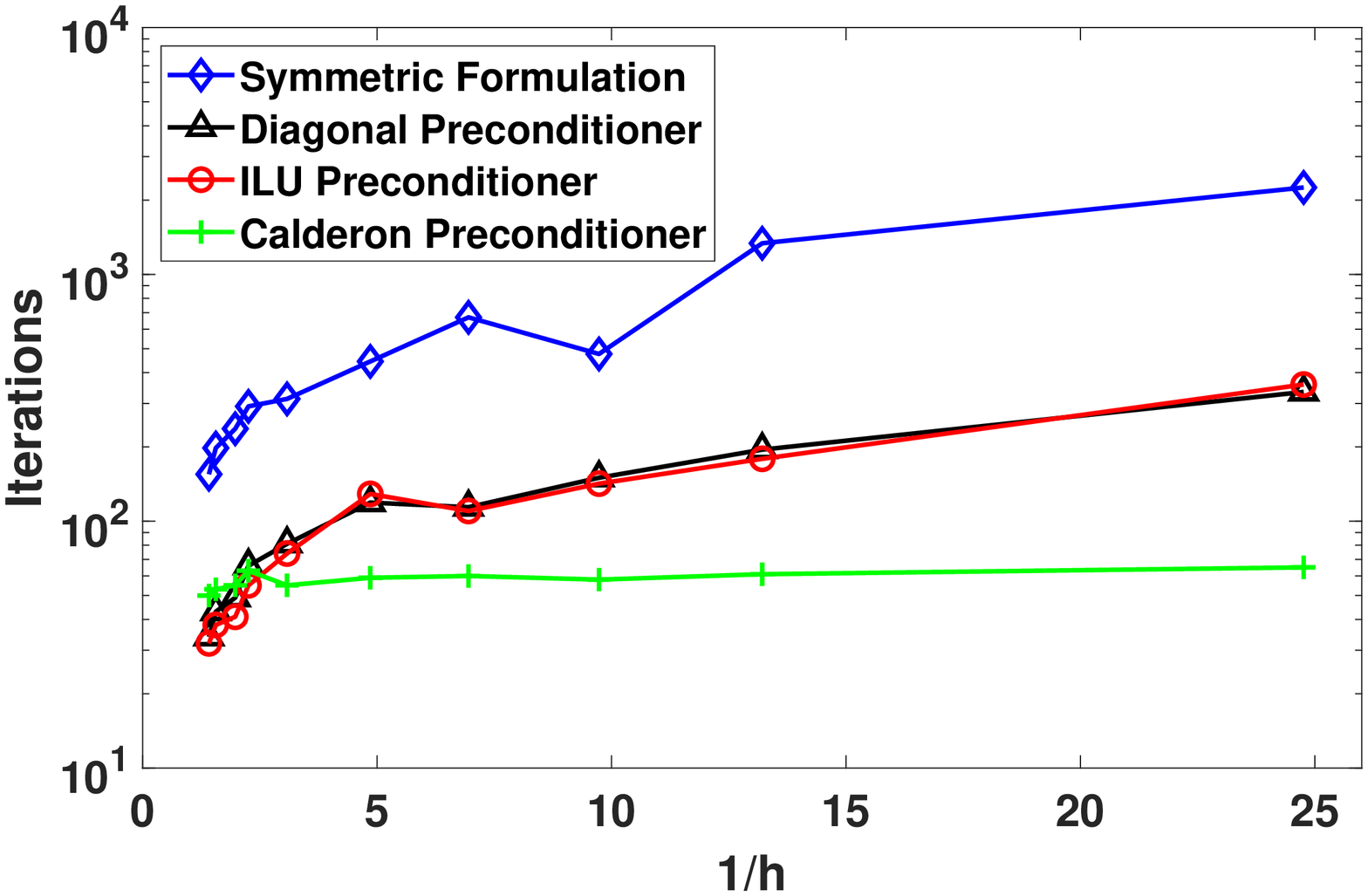}}
	\hspace{1pt}
	\subfloat[Number of iterations vs Conductivity Ratio\label{fig:iterRatio}]{\includegraphics[width=0.45\textwidth]{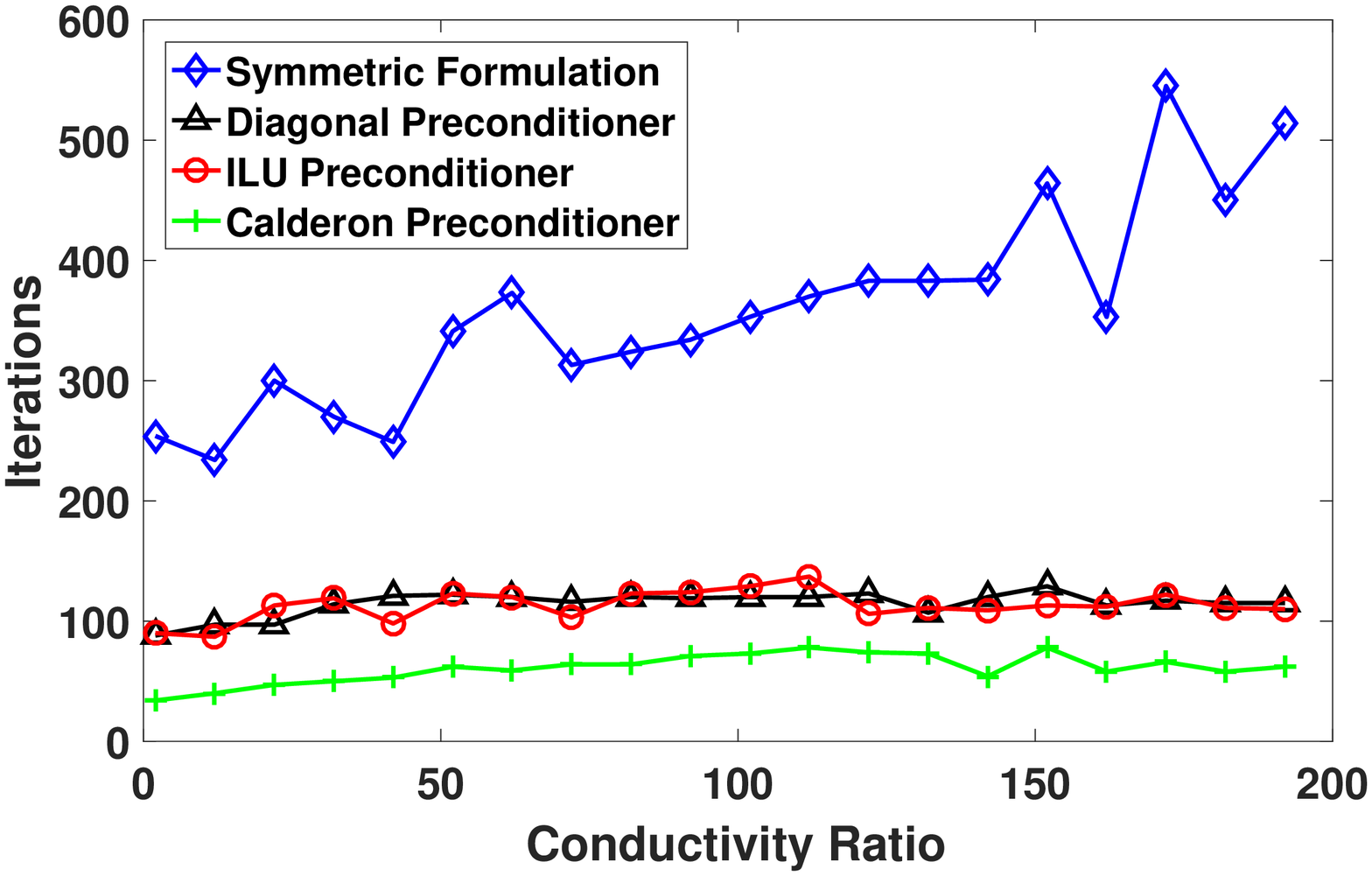}}
	\caption{Number of iterations for an accuracy of 1e-6 for different preconditioners.}
	\label{fig:Iter}
\end{figure}

The third test aimed at showing the efficiency in terms of number of iterations of the proposed preconditioner. For different spectral indices $1/h$, we present in Fig.~\ref{fig:iterEdge} the number of iterations needed for a Conjugate Gradient Square (CGS) solver to reach a relative residual error of $10^{-6}$. In this figure, it is clear that only with the proposed Calderon preconditioner this number of iterations stays constant with the mesh refinement. The number of iterations for increasing conductivity ratio is shown in Fig.~\ref{fig:iterRatio}. In this case, all the preconditioners lead to a constant number of iterations. Still, the best performance is given by the Calderon preconditioner.

\subsection{Assessments on time}

Since for small number of unknowns direct solvers can be used to solve the obtained system of equations, a preconditioner is only useful when the number of unknowns becomes too high for direct inversion, that is when it is necessary to use iterative solvers. Indeed, if $N$ denotes the number of unknowns, then the complexity of a direct solver is $O(N^3)$ while the complexity of an iterative solver is $O(k N^2)$, where $k$ is the number of iterations needed to reach a desired accuracy.
However, for high number of unknowns, given that BEM matrices are dense, the time needed to compute the system matrix is also increasing. To deal with this issue and to show the benefits of the proposed Calderon preconditioner in a high number of unknowns context we coupled it with an Adaptive Cross Approximation (ACA) algorithm \cite{zhao2005adaptive}, that provides a compressed version of the system matrix.

\begin{figure}[!h]
	\centering
	\includegraphics[width=0.55\textwidth]{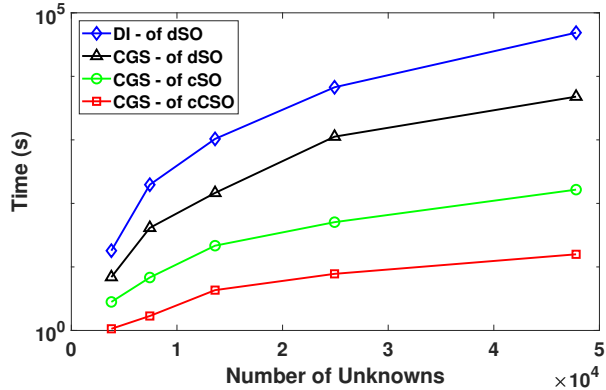}
	\caption{Time for solving one equation with respect to the mesh refinement. DI - Direct Inversion (DI), CGS - Conjugate Gradient Square, 
	dSO - dense Symmetric Operator, cSO - compressed Symmetric Operator, cCSO - compressed Calderon-Symmetric Operator}
	\label{figure:fmmTimeSol}
\end{figure}

\begin{figure}[!h]
	\centering
	\includegraphics[width=0.6\textwidth]{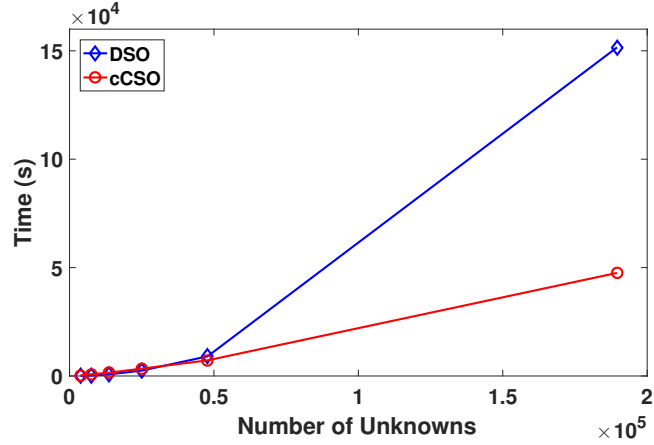}
	\caption{Time for computing the different Operators with respect to the mesh refinement. 
	DSO - Dense Symmetric Operator, cCSO - compressed Calderon-\textcolor{black}{Symmetric} Operator}
	\label{figure:fmmTimeOper}
\end{figure}

The time necessary to compute the solution for different numbers of unknown is shown Fig.\ref{figure:fmmTimeSol}. In this test, we compare the performance of direct and iterative solutions with or without the proposed Calderon preconditioner. It is clear that direct inversion (DI) has, as expected, a time complexity that increases rapidly with $N$. This prevents the use of this solver for very detailed models. The iterative \textcolor{black}{solver} used is \textcolor{black}{the} CGS algorithm. We see that without using any preconditioner the time necessary to obtain the solution with this iterative solver is also increasing with the number of unknowns even if this choice of solver is faster than DI. This is due to the fact the number of iterations increases with the mesh refinement parameter (due to the ill-conditioning of the matrices).  However, employing the proposed Calderon preconditioner solves this issue. \\
The time for computing the dense and compressed Calderon-Symmetric operator is presented in Fig.~\ref{figure:fmmTimeOper}. It can be seen that the use of the ACA yields in a linear time complexity

\subsection{Assessments on a MRI-obtained head model}

\begin{figure}[h!]
	\centering
	\hspace{5pt}
	\subfloat[Computed potential with a realistic mesh]{\label{fig:potScalp}\includegraphics[width=0.4\textwidth]{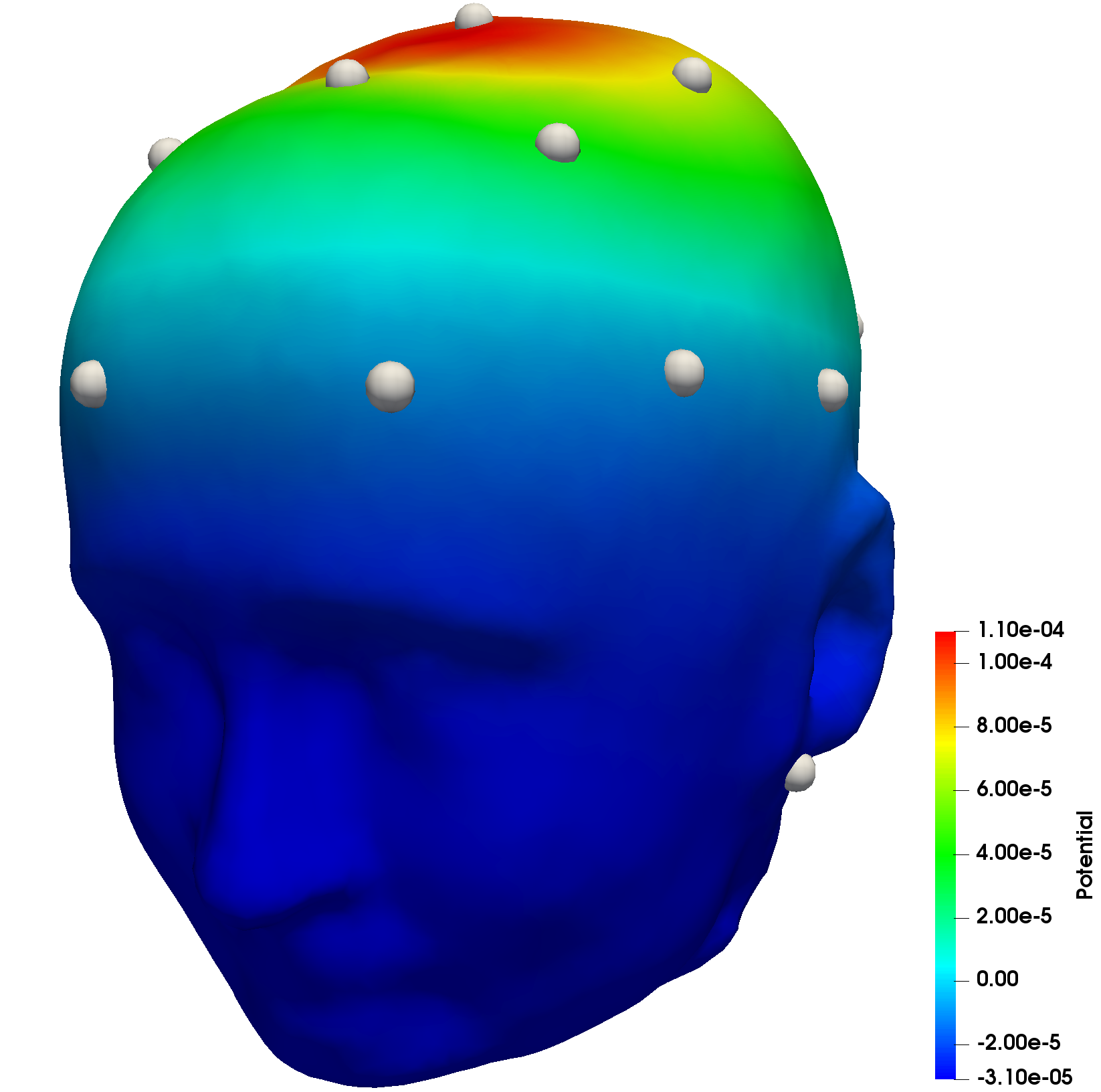}}
	\hspace{5pt}
	\subfloat[Error of the cCSO compared with DSO]{\label{fig:errorPot}\includegraphics[width=0.4\textwidth]{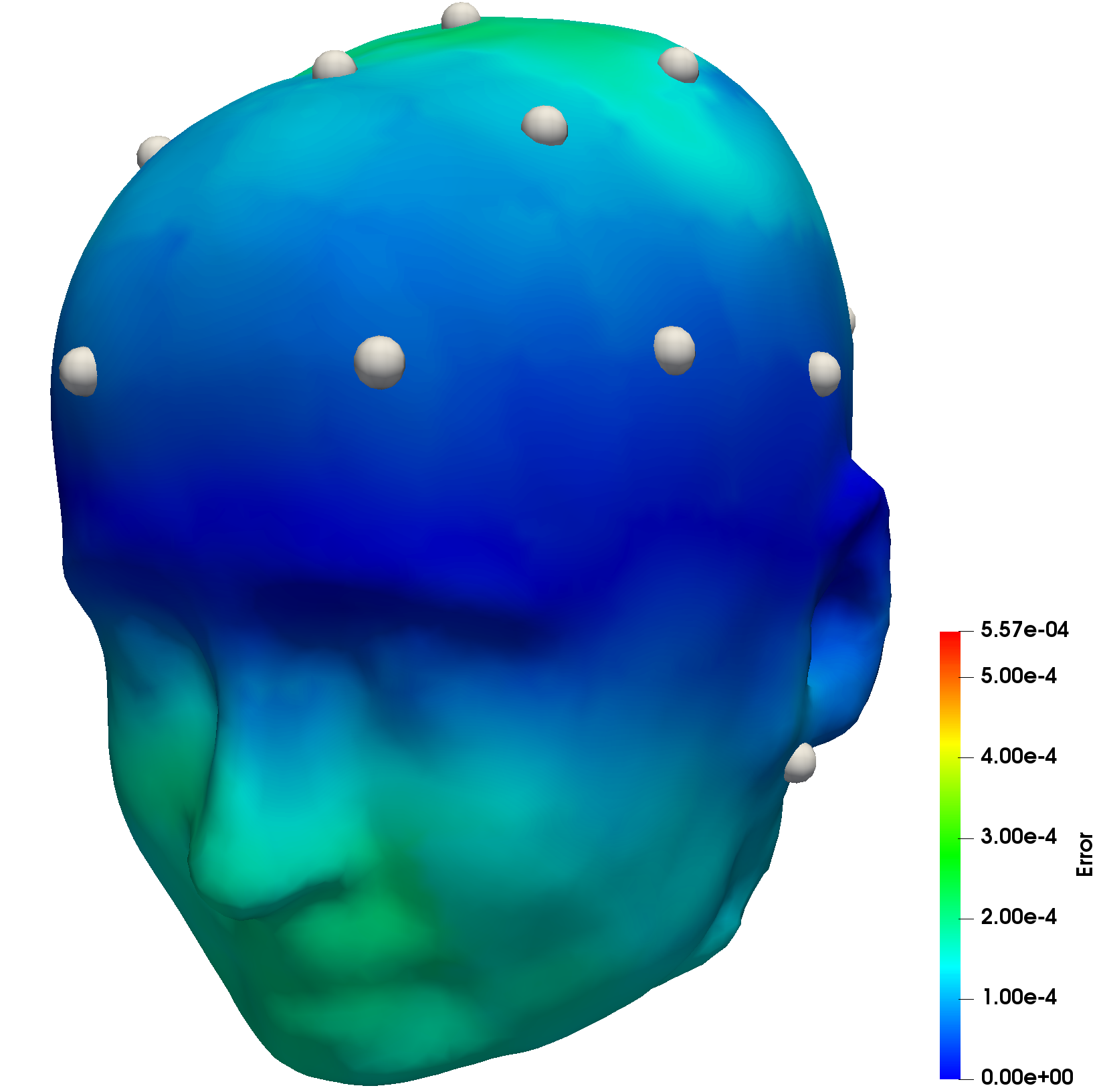}}
	\caption{MRI-obtained head model}
	\label{figure:MRI-obtained}
\end{figure}

\textcolor{black}{Finally, we seek to assess the performance of the proposed preconditioner in a realistic scenario that consists in computing a leadfield matrix using a head model obtained from MRI data. This matrix that provides the propagation model between known brain electric current sources and electrodes situated at the surface of the head of a patient is a key element in distributed inverse solution. For this purpose, we constructed a three layer mesh using \cite{oostenveld2010fieldtrip}. These layers model the brain, the skull, and the scalp. They contain 11850, 11616 and 22948 triangular cells respectively. The potential generated on the scalp by a single dipole situated in the brain is presented Fig.~\ref{fig:potScalp}. This figure also shows the position of the 21 electrodes for which we computed the leadfield matrix. The error when the Calderon-Symmetric formulation is solved with the ACA compared with the original \textcolor{black}{Symmetric} formulation solved with direct inversion is displayed Fig.~\ref{fig:errorPot}. It can be seen that the error is never greater than $0.05 \%$. To fill-in the leadfield matrix, we placed 1500 unitary dipole source in the brain layer, each having an orientation orthogonal to the brain surface. Using reciprocity  \cite{weinstein2000lead}, the forward model is then solved at each electrode position.
We compared in Table~\ref{table:LeadField}, for four different cases, the time needed to compute the operator, the time needed to solve the forward system once, and the time needed to compute the full leadfield matrix. Hence, the total time needed to get the leadfield matrix is given by the sum of the computation for obtaining the operator and the leadfield matrix. It can be seen that even if the time necessary to compute the compressed Calderon-Symmetric operator is greater than the compressed Symmetric operator, the fast convergence of the new method allows to compute the complete leadfield matrix in 2.56 hours, that is almost 10 times faster than without the proposed preconditioner. This compensates largely the computation overload in computing the preconditioning operator.}

\begin{table*}[t]
    \small
    \begin{center}
    \begin{tabular}{ | l | r | r | r | r |}
    \hline
    Method & Memory (GB) & Operator Time (s) & Solution Time (s) & Lead Field Time (h) \\ \hline
    DI Dense Symmetric & 16.234 & 10845.67 & 54609.99 & 18.19 \\ \hline
    CGS Dense Symmetric & 16.234 & 10845.67 & 7294.61 & 45.56 \\ \hline
    CGS Compressed Symmetric & 1.254 & 1436.63 & 2322.89 & 13.95 \\ \hline
    CGS Compressed Calderon-Symmetric & 2.542 & 7888.86 & 62.40 & 2.56 \\ 
    \hline
    \end{tabular}
    \caption{Memory and computation time information for computing a lead field matrix using the reciprocity method}
    \label{table:LeadField}
    \end{center}
    \normalsize
\end{table*}

\section{Conclusion}

\textcolor{black}{In this work, leveraging on Calderon identities, we have proposed a Calderon preconditioned symmetric formulation for the EEG forward problem. When compared to the standard symmetric equation, the proposed  formulation has the advantage of showing constant condition numbers both as a function of the mesh refinement and of the conductivity contrast. Employing this preconditioner does not degrade the accuracy of the symmetric formulation. This means that, for a given relative accuracy of the solution, the proposed formulation converges substantially faster than the standard one.
Moreover, the integration of the proposed approach into existing symmetric formulation implementations is achieved only at the cost of computing, on a barycentric refined mesh, the preconditioning operators with already existing tools.
Numerical results have substantiated the theoretical claims and have shown the practical impact of the newly proposed scheme.}

\section*{Acknowledgement}
This work was supported in part by CominLabs project SABRE under the reference ANR-10-LABX-07-01, in part by the ANSES project ECLAIR under the grant EST-2016-2 RF-23, and in part by the European Research Council (ERC) under the European Union’s Horizon 2020 research and innovation programme (grant agreement No 724846, project 321).



%\bibliographystyle{elsarticle-harv} 
% \bibliographystyle{elsarticle-num} 
\bibliographystyle{unsrt}
\bibliography{BiblioSymCald.bib}



%%\bibitem[ ()]{}

%%\end{thebibliography}
\end{document}